\documentclass{article}

\usepackage{arxiv}

\usepackage[utf8]{inputenc} 
\usepackage[T1]{fontenc}    
\usepackage[hidelinks]{hyperref}
\usepackage{url}            
\usepackage{booktabs}       
\usepackage{amsfonts}       
\usepackage{nicefrac}       
\usepackage{microtype}      
\usepackage{lipsum}		
\usepackage{graphicx}
\usepackage{natbib}
\usepackage{doi}
\usepackage{amsmath}
\usepackage{newtxtext}
\usepackage{newtxmath}
\usepackage{mathdots}
\usepackage{tikz}
\usetikzlibrary{shapes}

\usepackage{xcolor}
\definecolor{Orange}{rgb}{1,0.5,0}
\definecolor{Cyan}{cmyk}{1,0,0,0}
\definecolor{Red}{rgb}{1,0,0}
\definecolor{Blue}{rgb}{0,0,1}

\newcommand{\markerup}{\raisebox{0.5pt}{\tikz{\node[draw,scale=0.3,isosceles triangle,shape border rotate=90,black,fill=white](){};}}}

\newcommand{\markerdown}{\raisebox{0.5pt}{\tikz{\node[draw,scale=0.3,isosceles triangle,shape border rotate=270,black,fill=white](){};}}}

\newcommand{\markercirc}{\raisebox{0.5pt}{\tikz{\node[draw,scale=0.4,circle,black](){};}}}

\newcommand\e{\mathrm{e}}
\newcommand\im{\mathrm{i}}

\title{Cross-frequency amplification of perturbations in a laminar separation bubble using resolvent analysis}

\author{\href{https://orcid.org/0000-0001-7791-5426}{\includegraphics[scale=0.06]{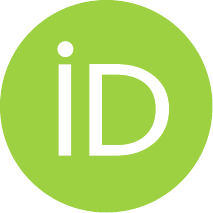}\hspace{1mm}
Md Rashidul Islam}\thanks{Corresponding author: mislam11@syr.edu}, \href{https://orcid.org/0000-0001-7808-672X}{\includegraphics[scale=0.06]{orcid.pdf}\hspace{1mm}Yiyang Sun} \\
	Department of Mechanical and Aerospace Engineering\\
	Syracuse University\\
	Syracuse, NY 13244, USA
}

\renewcommand{\shorttitle}{Cross-frequency amplification of perturbations in LSB}

\hypersetup{
pdfkeywords={First keyword, Second keyword, More},
colorlinks = {true},
urlcolor   = {black},
citecolor  = {blue},
}

\begin{document}
\maketitle

\begin{abstract}
A large-eddy simulation (LES) of a laminar separation bubble (LSB) induced by an adverse pressure gradient over a flat plate is performed at an inflow displacement-thickness-based Reynolds number of $410$ and a free-stream Mach number of $0.25$. With a mean peak reverse flow of $21.4\%$, the bubble sustains self-excited vortex shedding through a local region of absolute instability, in the absence of any external forcing. Spectral proper orthogonal decomposition (SPOD) applied to the LES data identifies three dominant coherent structures within the LSB: two-dimensional and oblique Kelvin--Helmholtz (KH) waves in the separated shear layer at the vortex-shedding frequency, and stationary spanwise-periodic streaks near reattachment at near-zero frequency. Classical resolvent analysis of the mean flow identifies strong convective amplification of the KH waves over a range of spanwise wavenumbers, but predicts only weak amplification in the low-frequency, streak-forming region, where the leading gain is orders of magnitude smaller and no dominant rank-one mechanism is present. This discrepancy with the SPOD energy indicates that the streaks are not sustained by same-frequency linear amplification, but are instead energized by the intrinsic forcing, which the classical framework treats as an unexplained input. Harmonic resolvent analysis of the time-periodic base flow reveals the underlying mechanism: the base-flow unsteadiness couples the oblique KH wave at the shedding frequency to the stationary streak through cross-frequency amplification, yielding a gain far larger than that of the direct same-frequency amplification. This cross-frequency route provides a likely explanation for how the stationary streaks observed near reattachment are energized.
\end{abstract}

\section{\label{sec:intro} Introduction}
The laminar boundary layer over a lifting surface often separates due to a sudden geometric change or a severe adverse pressure gradient (APG) along the flow direction. The separated shear layer, which is susceptible to environmental disturbances, can undergo a transition from laminar to turbulent flow, enhancing momentum mixing and reattaching to the surface if the geometry is sufficiently long. In such a scenario, a closed region of recirculating flow exists in the time-averaged field between the separation and reattachment locations, known as the laminar separation bubble \citep{tani1964low,gaster1967structure}. In this work, we focus on the APG-induced LSB, which is characteristic of aerodynamic bodies operating at low to moderate chord Reynolds numbers, where the boundary layer tends to separate while remaining laminar; examples include turbomachinery blades, unmanned-air-vehicle wings, and airfoils at high angles of attack near takeoff. The formation of the LSB on the aerodynamic surface can degrade performance by increasing drag and reducing lift, thereby diminishing efficiency. Recent studies of pitching airfoils \citep{visbal2018exploration,visbal2023passive} have also highlighted the role of the breakdown of an initially small LSB in producing an increasingly larger separation zone, which leads to the formation of a dynamic stall vortex. Understanding the physics of separated flow and exploring ways to manipulate the LSB are therefore important for improving the performance of such lifting surfaces in engineering applications.

The separated shear layer in the LSB amplifies background disturbances, whose growth drives the transition from laminar to turbulent flow. Understanding the stability characteristics and transition mechanism in the LSB can facilitate the development of methods to predict and control the separated flow region, motivating significant research efforts to uncover the instability of the airfoil LSB using modal analysis \citep{jones2008direct,yarusevych2017steady,yeh2020resolvent,kurelek2016coherent}. An APG-induced LSB on a flat plate exhibits the same fundamental physical properties as the airfoil LSB and has been used as a canonical simplified model in many studies to investigate stability and transition characteristics \citep{balzer2016numerical,hosseinverdi2019numerical,borgmann2025experimental,marxen2010mean,marxen2011effect}. Earlier studies on the local linear stability (LST) of the LSB treated the base flow upstream of separation as parallel to the wall, with variation only in the wall-normal direction. Assuming small perturbations on this parallel base flow, LST has been shown to accurately predict the most amplified frequency in the LSB in both numerical and experimental studies \citep{marxen2003combined,yeh2020resolvent,michelis2017response,borgmann2025experimental,marxen2010mean}. A consensus regarding the primary instability is that it is of the inviscid Kelvin--Helmholtz type, originating from the inflectional velocity profile of the separated shear layer. In the absence of external disturbances, the absolute instability of the KH wave over a local spatial region can produce self-excited vortex shedding once the peak reverse flow reaches $\approx 15$--$25\%$ within the bubble \citep{alam2000direct,diwan2009origin}; below this threshold, vortex shedding instead arises through the saturation of a KH wave amplified from upstream disturbances \citep{hosseinverdi2019numerical,michelis2018origin}. The deformation and breakdown of the spanwise-uniform (2D) shed vortices naturally render the flow three-dimensional, motivating global stability analyses of the 3D instability mechanisms within the bubble. A common finding across these studies \citep{rodriguez2010structural,theofilis2000origins,cherubini2010onset} is that a stationary 3D global mode of centrifugal origin can develop within the bubble, driving a steady three-dimensionalization of the flow. \cite{rodriguez2013two} proposed an alternative scenario in which this stationary global mode acts as the primary instability (at $\approx 7\%$ peak reverse flow), rendering the flow three-dimensional before the absolute instability sets in and self-excites the vortex shedding. Regardless of the route, transition to turbulence is followed by reattachment and the formation of the closed LSB \citep{marxen2011effect}. Because transition precedes reattachment, promoting or delaying it in the shear layer shifts the reattachment location and, in turn, the overall size of the bubble. Understanding the effects of both 2D and 3D external disturbances on transition in the separated shear layer therefore remains an active area of research.

This sensitivity of the separated shear layer to disturbances can be exploited for control: one way to improve the aerodynamic performance of lifting surfaces near stall is to leverage the LSB's amplification of external disturbances through active flow control (AFC) \citep{collis2004issues,esfahani2019flow,wu2018response}. The idea is that, by tuning the actuator to the frequencies most amplified by the LSB, a large control effect can be achieved with minimal energy input. Recent experimental efforts \citep{michelis2017response,borgmann2025active,kurelek2023superposition,toppings2026bursting,yarusevych2017effect} have applied this idea to the LSB over a flat plate and over static and pitching airfoils, showing that 2D periodic forcing upstream of separation is most effective at reducing the bubble extent when its frequency lies near the most amplified frequency of the unforced LSB. Numerical studies employing similar 2D forcing have likewise reported a reduction in LSB extent \citep{alam2000direct,marxen2011effect,embacher2014direct,marxen2003combined,visbal2018exploration}. The underlying mechanism is the amplification of the forced disturbance as a KH wave in the separated shear layer, producing enhanced spanwise-coherent vortex shedding that promotes reattachment and shrinks the separation zone. As the forcing amplitude increases, the KH wave saturates more rapidly, causing earlier vortex rollup and transition, and thus a shorter LSB \citep{marxen2010mean,marxen2011effect}. In contrast, a spanwise-periodic steady 3D disturbance of moderate initial amplitude ($<3\%$) was found to have no significant effect on transition \citep{marxen2004effect}. These studies indicate that understanding the perturbation-amplification characteristics of the LSB, together with the physical mechanisms underlying this amplification, is critical for guiding the design of AFC. Because exploring the large control-parameter space through experiments or high-fidelity simulations is prohibitively costly, a complementary, model-based question is which forcing the flow amplifies most, and at what frequency and spatial structure.

In recent years, resolvent analysis has become a valuable tool for modeling the forced linear amplification of perturbations in fluid flows \citep{mckeon2010critical,jovanovic2004modeling}. In the classical approach, the Navier–Stokes equations are linearized about a statistically stationary base flow and, in the frequency domain, recast in an input-output framework, with the terms containing perturbation nonlinearities replaced by an unknown forcing \citep{mckeon2010critical}. The dynamics of the input and output perturbations are governed by the resolvent operator, whose singular value decomposition (SVD) identifies the most amplified response to an optimal forcing at a single frequency, ranked by the singular values \citep{rolandi2024invitation}. The use of the time-averaged flow as the stationary base for linearization has demonstrated the capability of the classical resolvent analysis (CRA) to model perturbation dynamics in turbulent flow \citep{yeh2019resolvent,liu2021unsteady,sun2020resolvent,rolandi2025biglobal,cura2025linear,thakor2024responses,yeung2024high}. Owing to the natural connection between its input-output formulation and flow-control design, the CRA has proven helpful for identifying instability mechanisms and devising control strategies for separated flows over airfoils \citep{yeh2019resolvent,yeh2020resolvent,rolandi2025biglobal}. For a high-Reynolds-number airfoil flow, \cite{yeh2020resolvent} performed a biglobal CRA of the time-averaged LSB and found that the 2D KH instabilities of the shear layer were the most amplified response to upstream forcing, with no dominant 3D amplification mechanism identified. The practical value of the CRA is illustrated by \cite{visbal2023passive}, where a micro-cavity on the pressure side of an airfoil reduced the downstream separated flow when the disturbances it introduced fell within the frequency range of the 2D LSB instabilities. An unsteady base flow, however, can amplify perturbations at frequencies other than the forcing frequency, and analysis using a stationary base may therefore overlook effective amplification routes arising from cross-frequency energy transfer.

This stationary assumption of the base flow in the CRA precludes modeling interactions among temporal frequencies, which can be important in flows exhibiting multi-frequency phenomena \citep{linot2025extracting}. To capture such cross-frequency coupling, \cite{padovan2020analysis} extended the framework by treating the base flow as time-periodic, with a fundamental frequency corresponding to a dominant unsteady mechanism already present in the flow. This unsteady base flow yields a frequency-domain transfer function, the harmonic resolvent operator \citep{padovan2022analysis,islam2024identification}, which resolves interactions between perturbations at different frequencies through the unsteady harmonics of the base flow. The SVD of this operator identifies the optimal forcing and response of the unsteady flow across a set of frequencies rather than a single frequency. The harmonic resolvent analysis (HRA) has been applied to reveal cross-frequency interactions in compressible cavity flow \citep{islam2024identification}, to model vortex pairing in a jet \citep{padovan2022analysis}, and to understand wave interactions in jet flow \citep{farghadan2026wave}. While the HRA models perturbation amplification across multiple frequencies, interpreting the gain between input and output is less straightforward, since the forcing that drives amplification is typically not distributed across frequencies; in flow-control applications, for instance, actuation is usually applied at a single frequency in an open-loop configuration. In such cases, one is often more interested in the gain between input and output at the same frequency, which also permits direct comparison with more established techniques such as the CRA, which is based on a linear time-invariant (LTI) system. The present framework retains this capability via the mean resolvent operator \citep{leclercq2023mean}, which yields an optimal LTI response to a forcing at a given frequency, amplified by the time-periodic base flow.

In this work, we investigate the 2D and 3D mechanisms underlying the forced perturbation dynamics of an LSB generated over a flat plate under an imposed APG using biglobal resolvent analyses. By constructing resolvent operators using both a stationary and an unsteady base flow of the LSB, we examine the role of base-flow unsteadiness in amplifying perturbations. The setup for the LES of the LSB, the method for identifying coherent structures from the flow data using SPOD, and the formulation of the classical, mean, and harmonic resolvent analyses are presented in Sec.~\ref{sec:methods}. The mean and unsteady flow characteristics of the LSB obtained from the LES are discussed in Sec.~\ref{sec:LSBcharcteristic}, and the dominant coherent structures are identified using SPOD in Sec.~\ref{sec:SPOD}. Based on the results of resolvent analyses, we examine the dominant non-modal linear amplification mechanisms in the LSB in Sec.~\ref{sec:resolvent}. Conclusions are given in Sec.~\ref{sec:conclusion}, and additional supporting discussion is provided in the appendices.

\section{\label{sec:methods} Methods}
In this section, we outline the methods used in the current study. We begin with an introduction to the LSB flow conditions and the LES numerical setup. We then describe the SPOD methodology, which will serve as a data-driven approach to identifying coherent structures in the nonlinear flow. Lastly, we present the formulation of the biglobal resolvent analysis frameworks used in this work to model coherent structures in a physics-based way.

\subsection{Large-Eddy Simulation}
\begin{figure}
\centering
\includegraphics[width=0.95\textwidth]{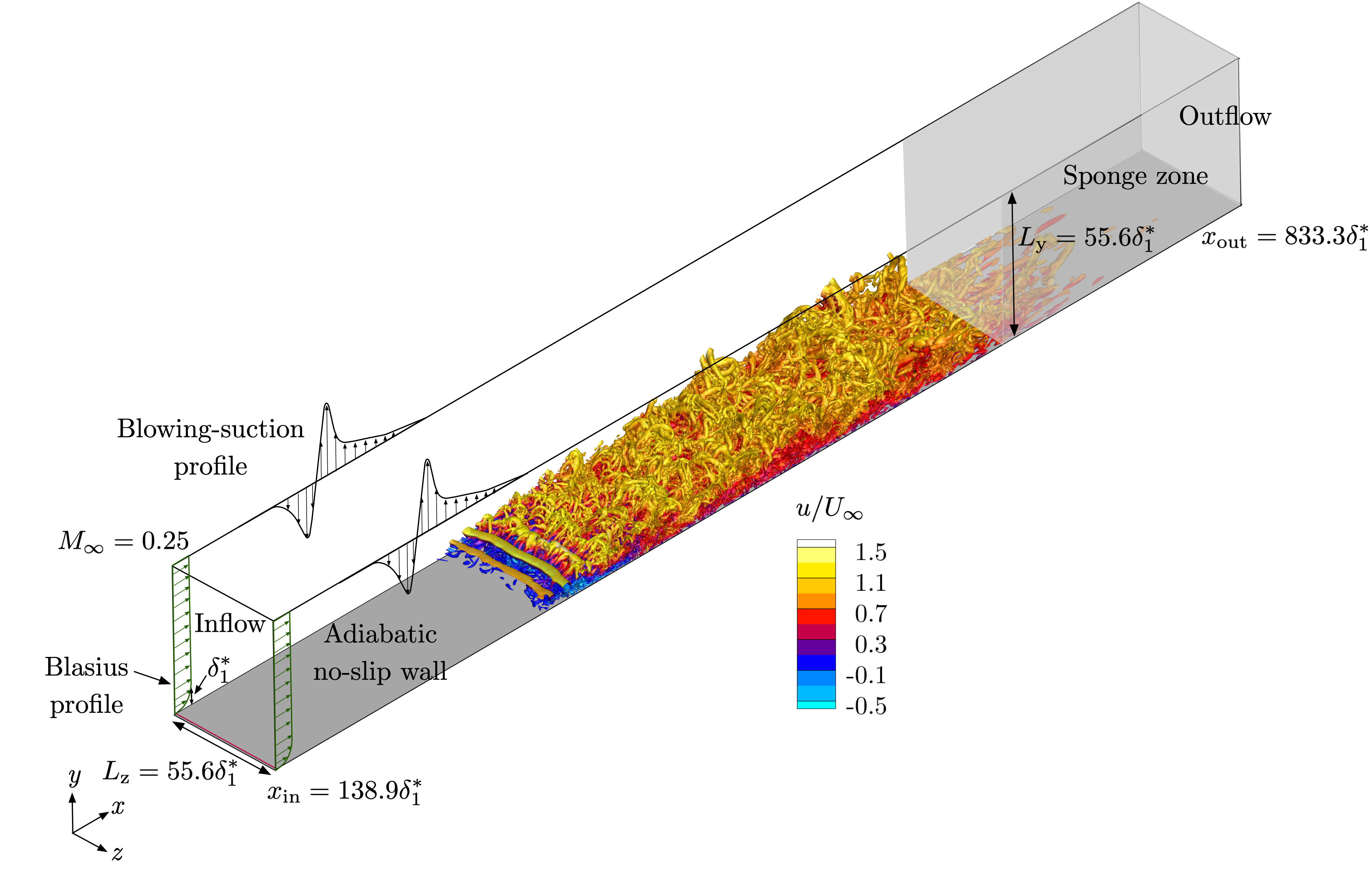}
\caption{\label{fig:1} Schematic of the computational domain used for the LES of an APG-induced LSB. Within the domain, an instantaneous 3D flow field is shown, visualized using the Q-criterion and colored by streamwise velocity.}
\end{figure}

We numerically study the flow of an LSB over a flat plate generated by an externally imposed streamwise pressure gradient as shown in Fig.~\ref{fig:1}. The numerical setup is guided by the experimental investigation of a similar flow by \cite{gaster1967structure}. In that experiment, the pressure gradient was imposed by placing an airfoil at a sufficiently large distance from the flat plate surface. However, due to computational expense, we exclude the displacement body from the numerical simulation domain, following earlier studies \citep{embacher2014direct,balzer2016numerical,marxen2011effect}. Rather, a velocity profile in the form of suction, blowing, or a combination of both can be applied at the top boundary to induce the same streamwise pressure gradient condition at the flat plate surface. In the current setup, we prescribe a boundary condition with streamwise-varying wall-normal velocity on the top surface of the domain, so that the resulting pressure gradient changes from favorable to adverse, as is typically observed in flow over an airfoil at a nonzero angle of attack. The simulation is performed at a free-stream Mach number of $M_{\infty} = 0.25$, corresponding to a weakly compressible flow. The global Reynolds number based on the inflow boundary layer displacement thickness $\delta^*_1$ and the free-stream velocity $U_{\infty}$ is set to $Re_{\delta^*_1} = U_{\infty} \delta^*_1/\nu = 410$, where $\nu$ is the kinematic viscosity. 

We perform 3D LES of the LSB flow over a flat plate using the solver \textit{CharLES} \citep{bres2017unstructured}. In \textit {CharLES}, the compressible Navier--Stokes equations are solved using a second-order finite volume method for the spatial discretization and a third-order Runge--Kutta temporal scheme with explicit time stepping. The LES is conducted using the Vreman subgrid-scale model \citep{vreman2004eddy}. The computational domain is shown in Fig.~\ref{fig:1}. The $x$, $y$, and $z$ coordinates are defined along the streamwise, wall-normal, and spanwise directions, respectively. The coordinate system origin is located upstream of the domain, at the virtual leading edge of the flat plate with a uniform incoming profile, and is centered in the spanwise direction. The computational domain lengths are made non-dimensional using the reference length $\delta_1^*$. The inlet of the domain is placed at a distance $x_{\text{in}}/\delta_1^* = 138.9$, and the outflow is located at $x_{\text{out}}/\delta_1^* = 833.3$. In the wall-normal direction, the domain length is $L_y/\delta_1^*=55.6$. As it is recommended to use a spanwise domain width of at least four times the maximum bubble height \citep{borgmann2025experimental} to capture the largest spanwise structures, we consider a spanwise domain extent of $L_z/\delta_1^*=55.6$, which corresponds to a spanwise width-to-maximum bubble height ratio of approximately 6.7 in the present setup.  

To obtain the wall-normal velocity profile specified at the top boundary, we perform an inviscid flow calculation based on Laplace's equation, assuming incompressible flow. The resulting profile is then used to compute the pressure gradient along the flat bottom plate. An iterative process is used to obtain the velocity profile such that the pressure gradient at the slip wall ($\boldsymbol{u}\cdot \boldsymbol{n}=0$) located at the flat plate, where $\boldsymbol{n}$ is a unit normal vector, matches a target inviscid pressure distribution identical to the experimental inviscid pressure distribution (Series I case 6) of \cite{gaster1967structure}. The details of the inviscid calculations are provided in Appendix~\ref{appA}. We apply a Neumann boundary condition for the streamwise velocity ($\partial u/\partial n = 0$) and specify the pressure using an isentropic relation at the top boundary. At the inflow boundary, we join a laminar Blasius velocity profile with displacement thickness $\delta^*_1$ within the boundary layer to the inviscid velocity from the potential flow solution outside the boundary layer to specify the streamwise velocity, along with the wall-normal velocity component of the potential flow. The Neumann condition for the pressure ($\partial p/\partial n = 0$) is applied at the inflow. The flat plate surface is treated as a no-slip, adiabatic wall. A sponge zone \citep{freund1997proposed} of length $L_{\text{sponge}}/\delta_1^*=222.2$ from the outflow boundary towards the inside of the domain is used to damp the outgoing waves. We apply periodic boundary conditions along the spanwise direction. The simulation is initialized by prescribing streamwise-varying Blasius laminar boundary-layer profiles in the whole domain. We do not apply any external disturbance, and the transition from laminar to turbulent flow occurs naturally.

The flow is resolved with $N_x\times N_y\times N_z=1400\times 128\times 128$ points in the streamwise, wall-normal, and spanwise directions, respectively, resulting in approximately 23 million grid points. In the streamwise direction, the grid spacing is uniform outside of the sponge zone. The grid in the wall-normal direction is stretched nonuniformly, with the spacing of the first grid point off the wall in viscous units set to $\Delta y^+=0.6$, which ensures the boundary layer is well resolved. The grid spacing is uniform in the spanwise direction. The resulting streamwise and spanwise spacing in viscous units in the attached flow regions are $\Delta x^+\approx10$ and $\Delta z^+\approx 8$, which are well within the requirements for performing wall-resolved LES.

\subsection{Spectral proper orthogonal decomposition}
To uncover the dominant coherent dynamics in the LES data, we perform the SPOD analysis \citep{lumley2012stochastic}. SPOD is a modal analysis technique that identifies spatially and temporally coherent structures in a time-resolved flow database by decomposing the cross-spectral density (CSD) matrix via an eigenvalue decomposition. To perform the SPOD, we follow the procedure outlined by \cite{towne2018spectral}, which we briefly describe below. 

The time series of flow snapshots, each with $p$ degrees of freedom, is divided into $N_b$ blocks, with $50\%$ overlap between samples. Then, data from each block is transformed into the frequency domain using the discrete Fourier transform (DFT), and the resulting Fourier coefficients at a particular angular frequency $\omega$ from all blocks are assembled into a data matrix
\begin{eqnarray}
\label{eq:spod1}
    \hat{\boldsymbol{Q}}(\omega) = \sqrt{\kappa} \begin{bmatrix}
        \hat{\boldsymbol{q}}_{\omega}^{(1)}& \hat{\boldsymbol{q}}_{\omega}^{(2)}& \hat{\boldsymbol{q}}_{\omega}^{(3)}& \dots& \hat{\boldsymbol{q}}_{\omega}^{(N_b)}
    \end{bmatrix} \in \mathbb{C}^{p\times N_b},
\end{eqnarray}
where $\kappa = \Delta t/(sN_b)$, $s$ is the sum of the squares of the window vector, and $\Delta t$ is the sampling time interval between snapshots. Before computing the DFT of each block, the block-wise temporal mean of the data is subtracted from the snapshots, and a Hamming window is applied to reduce the spectral leakage. The CSD matrix at each frequency is estimated as $\hat{\boldsymbol{S}}(\omega) = \hat{\boldsymbol{Q}}(\omega) \hat{\boldsymbol{Q}}^*(\omega) \in \mathbb{C}^{p\times p}$. The operator $(\cdot)^*$ denotes the Hermitian transpose. The SPOD eigenvalues $\boldsymbol{\Lambda}(\omega)$ and eigenvectors $\boldsymbol{\Psi}(\omega)$ can be computed as
\begin{eqnarray}
\label{eq:spod2}
    \hat{\boldsymbol{S}}(\omega) \boldsymbol{W} \hat{\boldsymbol{\Psi}}(\omega)  = \hat{\boldsymbol{\Psi}}(\omega) \boldsymbol{\Lambda}(\omega),
\end{eqnarray}
where $\boldsymbol{W}$ is a weighting matrix. The eigenvalues in $\boldsymbol{\Lambda}(\omega)$ represent the SPOD energy, and the SPOD modes, i.e., columns in $\hat{\boldsymbol{\Psi}}(\omega)$, are ranked by the energies. Since the degrees of freedom $p$ in the problem are much larger than the number of blocks $N_b$, we use the method of snapshots \citep{sirovich1987turbulence} to perform the SPOD in a computationally feasible way, following
\begin{eqnarray}
\label{eq:spod3}
    \hat{\boldsymbol{Q}}^*(\omega) \boldsymbol{W} \hat{\boldsymbol{Q}}(\omega) \tilde{\boldsymbol{\Psi}}(\omega)  = \tilde{\boldsymbol{\Psi}}(\omega) \boldsymbol{\Lambda}(\omega),\\
    \hat{\boldsymbol{\Psi}}(\omega) = \hat{\boldsymbol{Q}}(\omega) \tilde{\boldsymbol{\Psi}}(\omega) \hat{\boldsymbol{\Lambda}}(\omega)^{-1/2},
\end{eqnarray}
where the problem is reduced to a much smaller matrix of size $N_b \times N_b$. 

\subsection{Biglobal Resolvent analyses}
In the present work, we apply the resolvent analysis \citep{mckeon2010critical,padovan2020analysis,islam2024identification} to investigate the non-modal linear amplification of perturbations in the flow. We assume the perturbations are homogeneous in the spanwise direction and over time, while the base flow is 2D varying along the streamwise ($x$) and wall-normal ($y$) direction, which classifies the analysis as biglobal \citep{rolandi2024invitation,liu2021unsteady,sun2014numerical}. The resolvent analyses are performed by considering both a stationary and unsteady base flow, yielding different variants; the formulation of each will be discussed next.

We begin the description of resolvent analysis considering a time-periodic flow as the base state, which will result in the formulation for the \textit{harmonic resolvent} analysis (HRA) \citep{padovan2022analysis,islam2024identification}. By specifying $\boldsymbol q = [\rho, \rho u, \rho v, \rho w, \rho E]^T$ as the vector of state variables, we express the conservative form of the Navier--Stokes equation in a compact form
\begin{equation}
    \frac{\partial \boldsymbol{q}(x,y,z,t)}{\partial t} = \mathcal{N}(\boldsymbol{q}(x,y,z,t)).
    \label{NS}
\end{equation}
We then decompose the state variables as $\boldsymbol{q}(x,y,z,t) = \boldsymbol{Q}(x,y,t) + \boldsymbol{q}'(x,y,z,t)$, where $\boldsymbol{Q}(x,y,t)= \boldsymbol{Q}(x,y,t+T_0)$ is a time-periodic base flow with fundamental period of $T_0$, and $\boldsymbol{q}'(x,y,z,t)$ is a perturbation developing around the base flow. Substituting the decomposition into the Eq.~(\ref{NS}), we obtain
\begin{equation}
    \frac{\partial \boldsymbol{q}'(x,y,z,t)}{\partial t} = \boldsymbol{A}(x,y,\boldsymbol{Q},\frac{\partial}{\partial x},\frac{\partial}{\partial y},\frac{\partial}{\partial z},t) \boldsymbol{q}'(x,y,z,t) + \boldsymbol{f}'(x,y,z,t),
    \label{LNS}
\end{equation}
where $\boldsymbol{A}$ is the Jacobian of the Navier--Stokes equation evaluated at the base flow $\boldsymbol{Q}(x,y,t)$ and $\boldsymbol{f}'$ is an unknown forcing. The operator $\boldsymbol{A}$ is also time-periodic with the same period of $T_0$ as $\boldsymbol{Q}(t)$. We can express the operator $\boldsymbol{A}$ using a temporal Fourier series as $\boldsymbol{A}(t) = \sum_{k =-\infty}^{\infty} \hat{\boldsymbol{A}}_{k\omega_0} \e^{\im k\omega_0 t}$, where $k$ is an integer, and $\omega_0 = 2\pi /T_0$ is the fundamental frequency. For the perturbations, we seek solutions of the form 
\begin{eqnarray}
    \boldsymbol{q}'(x,y,z,t) = \e^{\im \gamma \omega_0 t} \sum_{j=-\infty}^{\infty} \hat{\boldsymbol{q}}'(x,y)_{\beta,(\gamma+j)\omega_0} \e^{\im (j \omega_0 t + \beta z)},\\
    \boldsymbol{f}'(x,y,z,t) = \e^{\im \gamma \omega_0 t} \sum_{j=-\infty}^{\infty} \hat{\boldsymbol{f}}'(x,y)_{\beta,(\gamma+j)\omega_0} \e^{\im (j \omega_0 t + \beta z)},
\end{eqnarray}
where $\beta = 2\pi/\lambda_z$ is the spanwise wavenumber of the perturbation, $j$ is an integer, and $\gamma\in[0,0.5]$ is a continuous parameter that sets the frequency offset from an integer multiple of $\omega_0$, while $j$ indexes the discrete harmonics. For $\gamma=0$, the perturbations have the same period $T_0$ as the base flow; whereas, for nonzero values of $\gamma$, we can study perturbations with a different period and corresponding frequency than the base flow. Substituting the Fourier series expansions in Eq.~(\ref{LNS}), we obtain
\begin{equation}
    \im [(\gamma+j) \omega_0] \hat{\boldsymbol{q}}'_{\beta,(\gamma+j) \omega_0} = \sum_{\substack{k \omega_0\in \Omega_{p}\\ (j-k)\omega_0 \in \Omega_b}} \hat{\boldsymbol{A}}(\beta)_{(j-k)\omega_0} \hat{\boldsymbol{q}}'_{\beta,(\gamma+k) \omega_0} + \hat{\boldsymbol{f}}'_{\beta,(\gamma+j) \omega_0},
    \label{HR1}
\end{equation}
for a particular value of $\gamma$ and $\beta$. The set of base flow frequencies $\Omega_b = \{\dots,-\omega_0,0,\omega_0,\dots\}$ and perturbation frequencies $\Omega_{\gamma}=\gamma\omega_0 +\Omega_p$ with $\Omega_{p} = \{\dots,\allowbreak -2\omega_0,\allowbreak -\omega_0,\allowbreak 0,\allowbreak \omega_0,\allowbreak 2\omega_0,\allowbreak \dots\}$ are infinite dimensional. The Eq.~(\ref{HR1}) represents a system of infinitely coupled equations where perturbation at the temporal frequency $(\gamma + j) \omega_0$ and spanwise wavenumber $\beta$ is coupled with perturbation at the temporal frequency $(\gamma + k) \omega_0$ and the same wavenumber $\beta$ through the 2D base flow at temporal frequency $(j - k)\omega_0$. However, in practice, the set of base flow frequencies $\Omega_b$ and the perturbation frequencies $\Omega_p$ are truncated to have finite numbers of harmonics $N_h$ and $N_p$, respectively, for a sufficient approximation. This results in $2N_h+1$ and $2N_p+1$ frequencies in the set $\Omega_b$ and $\Omega_p$, respectively. Using the Toeplitz transformation, the coupled set of equations in Eq.~(\ref{HR1}) can be represented in a matrix form as
\begin{equation}
\label{eqn10}
    [\mathrm{i}\gamma\omega_0 \boldsymbol{I} - \boldsymbol{T}(\beta)] \hat{\boldsymbol{\mathcal{Q}}}'_{\beta,\gamma} = \hat{{\boldsymbol{\mathcal{F}}}}'_{\beta,\gamma},
\end{equation}
where the operator $\boldsymbol{T}$ is given by
\begin{equation}
    \boldsymbol{T}(\beta) = 
    \begin{bmatrix}
        & \vdots& \vdots& \vdots& \\
        \dots&\im \omega_0 \boldsymbol{I} + \hat{\boldsymbol{A}}(\beta)_0& \hat{\boldsymbol{A}}(\beta)_{-\omega_0} & 0& \dots\\
        \dots&\hat{\boldsymbol{A}}(\beta)_{\omega_0}& \hat{\boldsymbol{A}}(\beta)_0& \hat{\boldsymbol{A}}(\beta)_{-\omega_0}& \dots\\
        \dots& 0 & \hat{\boldsymbol{A}}(\beta)_{\omega_0}& -\im \omega_0 \boldsymbol{I}+\hat{\boldsymbol{A}}(\beta)_0& \dots\\
        & \vdots& \vdots& \vdots& 
        \label{optT}
    \end{bmatrix},
\end{equation}
and 
\begin{align}
    \setlength{\arraycolsep}{4pt}
    \hat{\boldsymbol{\mathcal{Q}}}'_{\beta,\gamma} = \renewcommand{\arraystretch}{1.6} \left[ \begin{array}{ccccc}
         \dots& \hat{\boldsymbol{q}}'_{\beta,(\gamma-1)\omega_0}& \hat{\boldsymbol{q}}'_{\beta,\gamma\omega_0}& \hat{\boldsymbol{q}}'_{\beta,(\gamma+1)\omega_0}& \dots
    \end{array} \right]^T,\\[4pt]
    \setlength{\arraycolsep}{4pt}
    \hat{\boldsymbol{\mathcal{F}}}'_{\beta,\gamma} = \renewcommand{\arraystretch}{1.6} \left[ \begin{array}{ccccc}
         \dots& \hat{\boldsymbol{f}}'_{\beta,(\gamma-1)\omega_0}& \hat{\boldsymbol{f}}'_{\beta,\gamma\omega_0}& \hat{\boldsymbol{f}}'_{\beta,(\gamma+1)\omega_0}& \dots
    \end{array} \right]^T.
\end{align}
The harmonic resolvent operator is defined as $\boldsymbol{H}(\beta,\gamma)= [\im \gamma\omega_0 \boldsymbol{I} - \boldsymbol{T}(\beta)]^{-1}$ \citep{islam2024identification,padovan2022analysis}, and the operator acts as a transfer function that maps an input $\hat{\boldsymbol{\mathcal{F}}}'_{\beta,\gamma}$ with temporal frequencies in the set $\Omega_{\gamma}$ to the output $\hat{\boldsymbol{\mathcal{Q}}}'_{\beta,\gamma}$ at the same temporal frequencies following the relation
\begin{equation}
    \hat{\boldsymbol{\mathcal{Q}}}'_{\beta,\gamma} = \boldsymbol{H}(\beta,\gamma) \hat{\boldsymbol{\mathcal{F}}}'_{\beta,\gamma}.
\end{equation}
 
We seek the unit norm optimal input $\hat{\boldsymbol{\mathcal{F}}}'_{\beta,\gamma}$ that maximizes the gain, i.e., generates the most amplified response $\hat{\boldsymbol{\mathcal{Q}}}'_{\beta,\gamma}$ mapped by the operator $\boldsymbol{H}(\beta,\gamma)$. Here, we considered the energy amplification measure using Chu's energy norm \citep{chu1965energy}, which is widely used in compressible flows. The optimization problem can be formulated in terms of the SVD of the weighted harmonic resolvent operator \citep{islam2024identification} as 
\begin{equation}
    \boldsymbol{W}^{1/2} \boldsymbol{M}^{-1} \boldsymbol{H}(\beta,\gamma) \boldsymbol{M} \boldsymbol{W}^{-1/2} = \tilde{\boldsymbol{U}} \boldsymbol{\Sigma} \tilde{\boldsymbol{V}}^*,
    \label{eqn7}
\end{equation}
where $\boldsymbol{W}$ is a positive definite weight matrix containing the weights of both the discrete numerical quadrature and the coefficient of Chu's norm, and the matrix $\boldsymbol{M}$ converts primitive perturbation state variables to conservative state variables. The columns in the matrix $\boldsymbol{U}=\boldsymbol{W}^{-1/2}\tilde{\boldsymbol{U}}$ contain the response modes, and the columns in $\boldsymbol{V}=\boldsymbol{W}^{-1/2}\tilde{\boldsymbol{V}}$ are the forcing modes at the temporal frequencies corresponding to the set $\Omega_{\gamma}$, ranked in a descending order by the singular values along the diagonal in $\boldsymbol{\Sigma}$. More details on the method can be found in our previous work \citep{islam2024identification,islam2025effect}. 

The concept of input and output at multiple frequencies, with a gain between them, may seem counterintuitive when the goal is to model single-frequency forcing that produces a response at the same frequency, as in many flow-control applications. An approximate LTI representation of single-frequency input-output dynamics can be obtained through the harmonic resolvent operator $\boldsymbol{H}(\beta,\gamma)$ by right-multiplication with a frequency prolongation operator $\boldsymbol{P}_f=\text{diag}(\boldsymbol{0},\dots,\boldsymbol{I},\dots,\boldsymbol{0})\in \mathbb{R}^{[(2N_p+1)\times 5n]\times 5n}$ and left-multiplication with a frequency restriction operator $\boldsymbol{P}_y^T$ where $\boldsymbol{P}_y=\text{diag}(\boldsymbol{0},\dots,\boldsymbol{I},\dots,\boldsymbol{0})\in \mathbb{R}^{[(2N_p+1)\times 5n]\times 5n}$ and $n$ is the number of discrete grid points. This operation extracts the sub-block $\boldsymbol{H}_{0,0}(\beta,\gamma)$ from the main diagonal of $\boldsymbol{H}(\beta,\gamma)$ that maps the input $\hat{\boldsymbol{f}}'_{\beta,\gamma\omega_0}$ at the center frequency of $\hat{\boldsymbol{\mathcal{F}}}'_{\beta,\gamma}$ to the output $\hat{\boldsymbol{q}}'_{\beta,\gamma\omega_0}$ at the same frequency. The operator $\boldsymbol{H}_{0,0}(\beta,\gamma)=\boldsymbol{P}_y^T\boldsymbol{H}(\beta,\gamma)\boldsymbol{P}_f$ is also known as the \textit{mean resolvent} operator \citep{leclercq2023mean}, which provides the optimal LTI approximation of the input-output dynamics of a time-varying base flow in the statistically steady regime. The mean resolvent analysis (MRA) involves performing an SVD of the mean resolvent operator, yielding the optimal forcing and response at the frequency $\gamma\omega_0$ and the spanwise wavenumber $\beta$. 
 
We also performed the \textit{classical resolvent} analysis by linearizing the dynamics in Eq.~(\ref{LNS}) about the spanwise-time-averaged base flow $\overline{\boldsymbol{Q}}$, to serve as a reference for how the non-modal amplification is modified by the unsteady base flow. The classical resolvent operator $[\im \omega \boldsymbol{I}-\boldsymbol{A}(\overline{\boldsymbol{Q}},\beta)]^{-1}$ \citep{mckeon2010critical} acts as a transfer function from an input $\hat{\boldsymbol{f}}'_{\beta,\omega}$ to the output perturbation $\hat{\boldsymbol{q}}'_{\beta,\omega}$ at the same single frequency, and its SVD yields pairs of forcing and response modes ranked by singular values. Although both the CRA and MRA yield optimal LTI responses, we emphasize that the fundamental difference between them lies in how the base flow is treated: in the CRA the base flow is stationary, whereas in the MRA it is time-periodic, allowing the perturbation to interact with the unsteady component of the base flow. The approximate LTI response to a deterministic forcing $\hat{\boldsymbol{f}}'_{\beta,\gamma\omega_0}$ predicted by the mean resolvent operator represents the phase-average of the responses to that forcing initialized at different phases of the periodic base flow \citep{leclercq2023mean}. 

The resolvent analyses are performed on a truncated domain spanning $x/\delta_1^*\in[138.9,638.9]$ in the streamwise direction, with a wall-normal domain length of $L_y/\delta_1^*=50$. At the flat plate surface, we specified the velocity perturbation ($\hat{\boldsymbol{u}}'$) and the wall-normal gradient of pressure perturbation ($\partial \hat{p}'/\partial n$) to zero with an adiabatic condition for the temperature perturbation ($\hat{T}'$). Sponge layers are applied within the regions $555.5<x/\delta_1^*<638.9$ near the outflow, $138.9<x/\delta_1^*<166.7$ near the inflow, and $33.3<y/\delta_1^*<50.0$ near the top boundary to bring the perturbation values to zero. The domain is discretized with $N_x\times N_y = 720\times 75$ grid points, with uniform spacing in the streamwise direction. The grid is stretched non-uniformly in the wall-normal direction from the smallest spacing adjacent to the flat plate surface to the largest at the top boundary. 

The SVD of the resolvent operators and the eigenvalues of the linear operators are computed using an in-house solver based on the PETSc \citep{balay2019petsc} and SLEPc \citep{hernandez2005slepc} libraries. The linear equations in the algorithm are solved directly based on an LU decomposition of the operator using the library MUMPS \citep{amestoy2000mumps}. The code used to generate the linear operator for the CRA has been validated in our earlier studies \citep{sun2017biglobal,sun2017global}. The linear operators for HRA are generated using a code validated against the airfoil flow results of \cite{padovan2020analysis}, with details provided in \citep{islam2024identification}. We have used $10$ random test vectors for the SVD calculation using the randomized algorithm, which we found sufficient to obtain converged results.

\section{\label{sec:LSBcharcteristic} Flow characteristics of the LSB}

\subsection{Mean flow characteristics}
\begin{figure}
\centering
\includegraphics[width=0.95\textwidth]{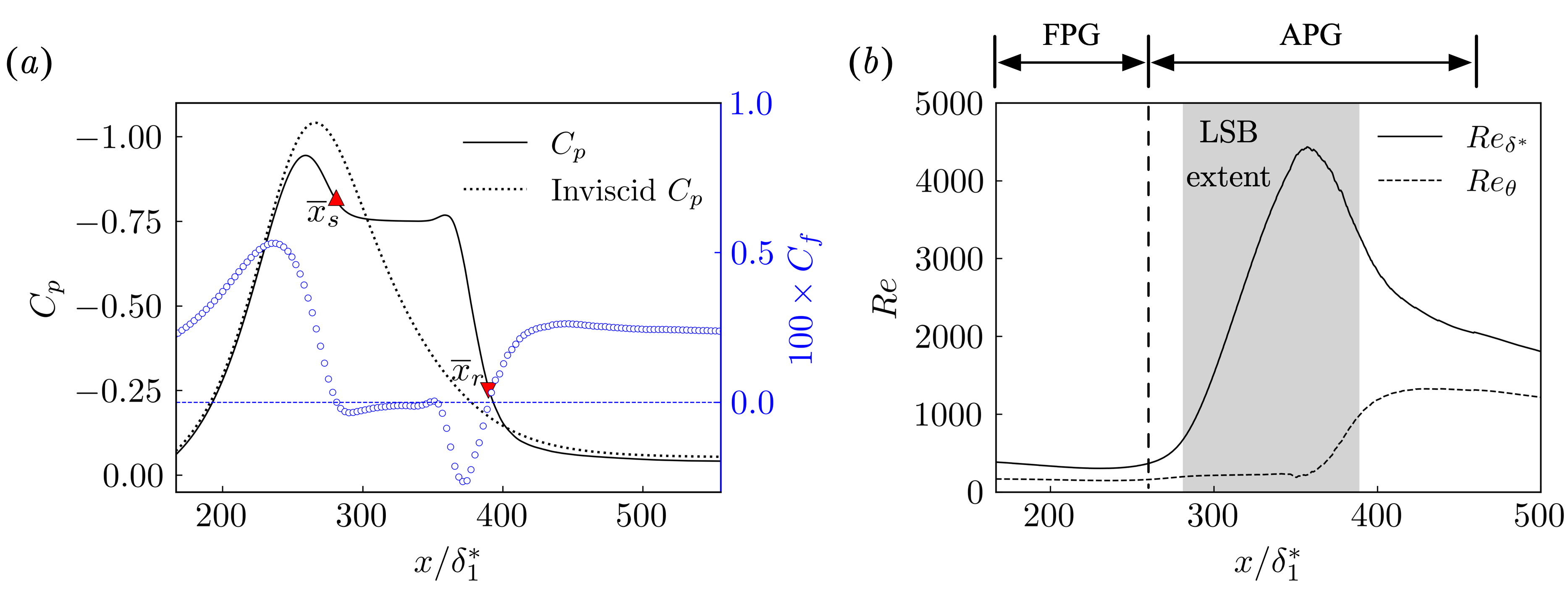}
\caption{\label{fig:3}(a) Time- and spanwise-averaged distribution of the pressure coefficient $C_p$ and the skin friction coefficient $C_f$ at the surface of the flat plate. (b) Streamwise variations of the Reynolds numbers based on the boundary layer displacement thickness $Re_{\delta^*}$ and the momentum thickness $Re_{\theta}$.}
\end{figure}

The pressure variation imposed at the flat-plate surface is shown using the spanwise- and time-averaged pressure coefficient $C_p = (\overline{p}-p_{\infty})/0.5\rho_{\infty} U_{\infty}^2$ obtained from the LES in Fig.~\ref{fig:3}(a), to identify the streamwise extent of the LSB formation. The inviscid pressure coefficient distribution obtained from the potential-flow solution along the flat-plate location is also plotted. In the viscous flow case, the presence of a mean separation bubble can be inferred from a plateau in the deviation of the LES pressure coefficient from the inviscid curve. The markers denoting the location of the mean flow separation $\overline{x}_s$ and reattachment $\overline{x}_r$ are estimated from the first and last zero crossing, i.e., sign changing from positive to negative and vice versa, of the mean skin friction coefficient $C_f=\tau_w/0.5\rho_{\infty} U_{\infty}^2$ along the streamwise direction, where $\tau_w$ is the wall shear stress. The respective values of the separation and reattachment are $\overline{x}_s/\delta_1^{*}=281.4$ and $\overline{x}_r/\delta_1^{*}=389.7$, resulting in a length of the mean separation bubble of $\overline{l}_b/\delta_1^{*}=108.3$. Here, we note that the mean skin friction coefficient changes sign over a short interval, $x/\delta_1^{*} \in[347.4,352.8]$, due to a small secondary recirculation bubble bounded within the primary recirculation zone.

Due to the presence of strong FPG and APG, determining the boundary-layer edge and the integral boundary-layer quantities is not straightforward since the free-stream velocity is not uniform outside the boundary layer. Here, we defined the boundary layer quantities using a pseudo velocity $\overline{u}_{\text{ps}}$ obtained by wall-normal integration of the spanwise- and time-averaged spanwise vorticity $\overline{\omega}_z$ \citep{spalart2000mechanisms,marxen2011effect}. The displacement thickness $\delta^*$ and the momentum thickness $\theta$ are then defined as 
\begin{eqnarray}
    \label{eq:3p1}
    \delta^* = \int_0^{y_{\text{max}}} \left(1-\frac{\overline{u}_{\text{ps}}(y')}{\overline{u}_{\text{ps}}(y_{\text{max}})}\right) \mathrm{d}y',\\
    \theta = \int_0^{y_{\text{max}}} \frac{\overline{u}_{\text{ps}}(y')}{\overline{u}_{\text{ps}}(y_{\text{max}})} \left(1-\frac{\overline{u}_{\text{ps}}(y')}{\overline{u}_{\text{ps}}(y_{\text{max}})}\right) \mathrm{d}y',
\end{eqnarray}
with 
\begin{eqnarray}
    \overline{u}_{\text{ps}}(y) = \int_0^{y} \overline{\omega}_z(y') \mathrm{d}y',
\end{eqnarray}
where the maximum limit of integration $y_{\text{max}}$ can be taken at an inviscid region of negligible vorticity. Here, we have set $y_{\text{max}}=22.2\delta_1^*$. The boundary-layer edge is defined as the wall-normal location ($y$) where the pseudo-velocity reaches $99\%$ of the local free-stream velocity $\overline{u}_{\text{ps}}(y_{\text{max}})$. The corresponding mean streamwise velocity ($\overline{u}$) at that wall-normal location is used as the local boundary-layer edge velocity $\overline{u}_e$. 

\begin{figure}
\centering
\includegraphics[width=0.95\textwidth]{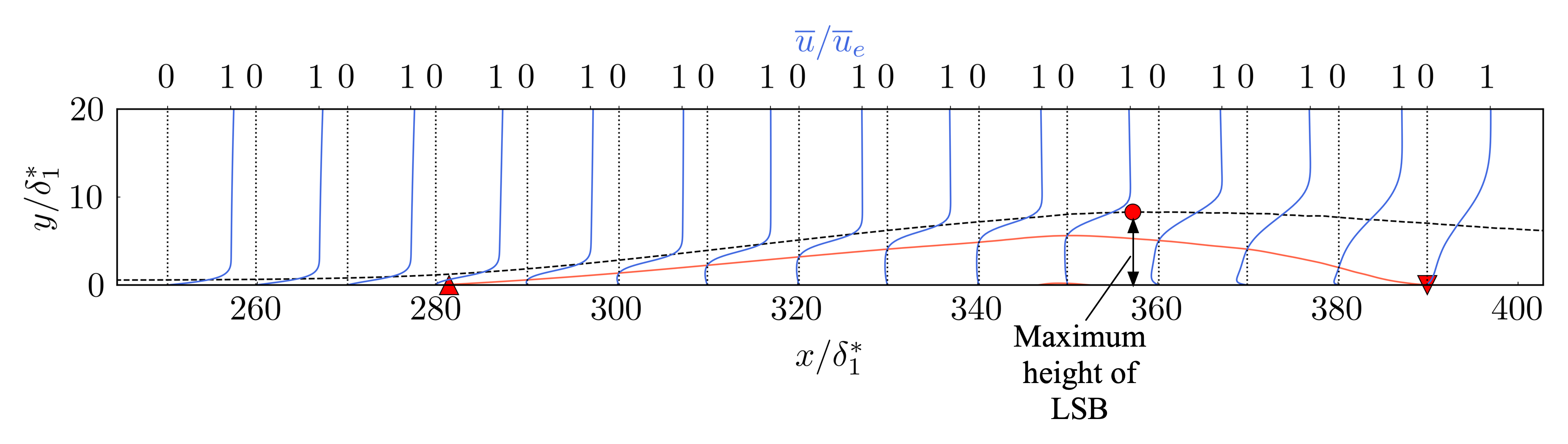}
\caption{\label{fig:4} Time- and spanwise-averaged streamwise velocity profiles normalized by the respective local boundary layer edge velocity. The solid red line depicts the boundary of the mean recirculation region using an iso-contour at $\overline{u}=0$. The dashed line denotes the boundary layer displacement thickness. \protect\markerup: separation, \protect\markerdown: reattachment, \protect\markercirc: maximum displacement thickness locations.}
\end{figure}

In Fig.~\ref{fig:3}(b), the Reynolds number variations along the streamwise direction are displayed using the newly defined boundary layer variables; i.e., $\theta,\delta^*,\overline{u}_{\text{e}}$; as $Re_{\theta}=\overline{u}_e\theta/\nu$ and $Re_{\delta^*}=\overline{u}_e\delta^*/\nu$. The displacement thickness in Fig.~\ref{fig:3}(b) in the FPG region decreases slightly due to the flow acceleration effect in the streamwise direction. In the APG region, the displacement thickness increases gradually, then rises abruptly at flow separation and peaks before reattachment. We have defined the height of the mean separation bubble ($\overline{h}_b$) to be the maximum value of the displacement thickness within the streamwise extent of the LSB, resulting in $\overline{h}_b/\delta_1^*=8.3$, located at $\overline{x}_{h_{\text{max}}}/\delta_1^*=357.2$, as shown in Fig.~\ref{fig:4}. After the bubble reaches its maximum height, the displacement thickness decreases with increasing streamwise distance toward the reattachment zone and continues to decrease downstream. In contrast, the momentum thickness increases slowly in the APG region prior to the maximum bubble height, then rises abruptly before reattachment. As we will see later in the instantaneous flow field, the roll-up of the shear-layer vortices occurs near the maximum bubble height, and their breakdown before reattachment enhances momentum mixing and wall shear, causing an abrupt increase in the momentum thickness. According to \cite{gaster1967structure}, the Reynolds number $Re_{\theta_s} = \overline{u}_{e,s}\theta_s/\nu$ based on the momentum thickness ($\theta_s$) and edge velocity ($\overline{u}_{e,s}$) at separation, and the pressure gradient parameter $P=\theta_s^2/\nu(\Delta u/\Delta x)$, where $\Delta u/\Delta x$ is the rise of inviscid velocity over the length of the bubble, are two important parameters that control the characteristics of the separation bubble. For the present LSB, the corresponding parameter values are $Re_{\theta_s}=198$ and $P=-0.161$, which classify the LSB as short \citep{gaster1967structure}.

The spanwise- and time-averaged profiles of the streamwise velocity ($\overline{u}$) illustrating the flow development along the streamwise direction are shown in Fig.~\ref{fig:4}. The velocity profiles are normalized by the local boundary-layer edge velocity $\overline{u}_e$ at each streamwise location. The iso-contour of the mean streamwise velocity corresponding to $\overline{u}=0$ is also plotted to delineate the boundary of the reverse-flow region. The peak value of the reverse flow ($\overline{u}_{\text{rev}}$) encountered within the separation bubble is approximately $16.9\%$ of the local boundary layer edge velocity ($\overline{u}_e$) and $21.4\%$ of the free-stream velocity ($U_{\infty}$). The peak reverse flow ($\overline{u}_{\text{rev}}$) occurs in the aft portion of the bubble, downstream of the streamwise location of the maximum displacement thickness, where vortex shedding develops. Since the mean peak reverse flow in the present LSB exceeds the $\approx 16\%$ of free-stream velocity level associated with the onset of absolute instability \citep{alam2000direct,rist2002investigations,diwan2009origin}, the self-excited vortex shedding observed here is likely sustained by a local region of absolute instability of the KH wave within the bubble.

\subsection{Unsteady flow characteristics}
\begin{figure}
\centering
\includegraphics[width=0.95\textwidth]{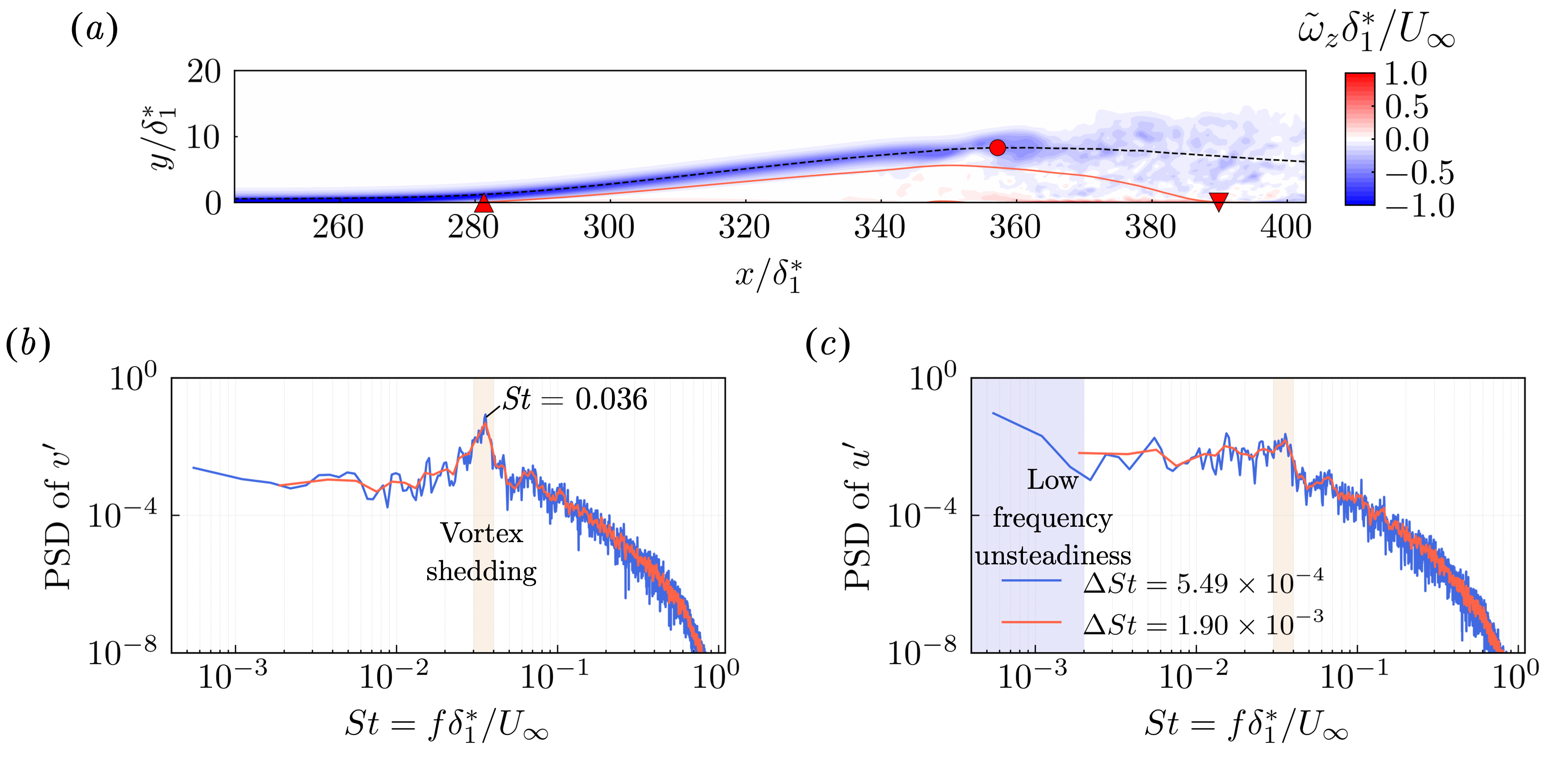}
\caption{\label{fig:5}(a) Contours of the instantaneous spanwise-averaged spanwise vorticity. Solid line: iso-contour at $\overline{u}=0$, dashed line: displacement thickness $\delta^*$, \protect\markerup: separation, \protect\markerdown: reattachment, \protect\markercirc: maximum displacement thickness locations. Power spectral density of the fluctuating (b) wall-normal ($v'$) and (c) streamwise velocity ($u'$), sampled at $(x,y,z)/\delta_1^*=(359.7,8.06,0.0)$.}
\end{figure}

To understand the dynamical behavior of the LSB, we show a representative instantaneous snapshot of the spanwise-averaged spanwise vorticity $\tilde{\omega}_z\delta_1^*/U_{\infty}$ in Fig.~\ref{fig:5}(a). The roll-up of vortices in the separated shear layer begins before the bubble reaches its maximum height $\overline{h}_b$ (indicated by a red circle), forming a spanwise-uniform roller that sheds near that height $\overline{h}_b$ and breaks down after traveling a short downstream distance. Reattachment occurs shortly after breakdown due to enhanced momentum exchange associated with the transition to turbulence. However, a fully developed turbulent flow is not expected in the present setup, which requires a longer streamwise distance. The spectral content of the wall-normal ($v'$) and streamwise ($u'$) velocity fluctuations sampled at the location $(x,y,z)/\delta_1^*=(359.7,8.06,0.0)$ close to the maximum bubble height is shown in Fig.~\ref{fig:5}(b) and (c), respectively. In addition to capturing the vortex shedding frequency, we also anticipate the possible presence of the so-called `flapping' phenomenon associated with the global motion of the separated shear layer, which typically occurs at temporal frequencies that are orders of magnitude lower than the vortex-shedding frequency \citep{zaman1989natural,michelis2018origin,malmir2024low}. Thus, the PSD of the fluctuating signals is estimated using the Welch method \citep{welch1967use} by dividing the temporal signals into $N_b=3$ and $N_b=12$ block segments, resulting in frequency resolutions of $\Delta St=5.49\times 10^{-4}$ and $1.90\times 10^{-3}$, respectively. The choice of block number is a compromise between resolving low frequencies (i.e., $N_b=3$), where a significant concentration of spectral energy is expected in the presence of flapping, and obtaining sufficient realizations for convergence (i.e., $N_b=12$). A $50\%$ overlap between the samples in the consecutive blocks, along with a Hamming window, is used. We note that the Strouhal number is defined as $St = f\delta_1^*/U_{\infty}$ to report the non-dimensional frequency in the present study. However, to compare some notable frequencies with the past studies in literature, we also report the corresponding normalized frequencies in terms of the Strouhal numbers $St_{\delta_s^*}=f\delta_s^*/\overline{u}_{e,s}$ and $St_{\theta_s}=f\theta_s/\overline{u}_{e,s}$ of the present work, where $\delta_s^*$ is the displacement thickness, $\theta_s$ is the momentum thickness, and $\overline{u}_{e,s}$ is the boundary layer edge velocity all measured at the streamwise location of separation ($\overline{x}_s$).

From the PSD of the wall-normal velocity fluctuations $v'$ in Fig.~\ref{fig:5}(b), we observe a well-defined peak around $St=0.036$, which is associated with the vortex shedding in the aft portion of the bubble. We compare this vortex shedding frequency with previous studies in terms of the Strouhal numbers $St_{\delta_s^*}$ and $St_{\theta_s}$ in Table~\ref{tab:table1}. The corresponding vortex shedding frequency of $St_{\delta_s^*}=0.033$ is comparable to the reported value of $0.038$ by \cite{michelis2017response}. The resulting vortex shedding frequency based on the momentum-thickness-based scaling $St_{\theta_s}=0.0095$ is also within the range of values $0.0089-0.012$ reported by previous LSB studies on a flat plate \citep{balzer2016numerical,rodriguez2021self,michelis2017response}. The PSD of the streamwise velocity fluctuations $u'$ obtained using the smaller frequency resolution in Fig.~\ref{fig:5}(c) shows an elevated energy content for frequencies lower than $St=0.002$. In the experimental study by \cite{michelis2017response}, an increased energy in the PSD of $u'$ was reported for $St_{\delta_s^*}<0.005$ due to a low-frequency flapping of the unforced LSB. The range of frequencies with the increased spectral energy in Fig.~\ref{fig:5}(c) corresponds to $St_{\delta_s^*} < 0.0018$, suggesting the presence of a flapping motion in the current LSB. 

\begin{table}[t]
\centering
\caption{\label{tab:table1}
Comparison of characteristic LSB parameters between the present result and earlier studies. The Reynolds number $Re_{\theta_s} = \overline{u}_{e,s}\theta_s/\nu$ is based on the momentum thickness $\theta_s$ and boundary-layer edge velocity $\overline{u}_{e,s}$ at separation ($\overline{x}_s$). The vortex shedding frequency is reported as the Strouhal numbers $St_{\theta_s}=f\theta_s/\overline{u}_{e,s}$ (based on $\theta_s$), $St_{\delta_s^*}=f\delta_s^*/\overline{u}_{e,s}$ (based on displacement thickness $\delta_s^*$ at separation), and $St=f\delta_1^*/U_{\infty}$ (based on the inflow displacement thickness $\delta_1^*$). The last column gives the peak reverse flow $\overline{u}_{\text{rev}}$ scaled by the local boundary-layer edge velocity $\overline{u}_{e}$.}
\begin{tabular}{cccccc}
 &$Re_{\theta_s}$ &$St_{\theta_s}$ &$St_{\delta_s^*}$& $St$& $\overline{u}_{\text{rev}}/\overline{u}_{e} (\times 100\%)$\\
\hline
\textrm{\cite{michelis2017response} - Experimental}& 345 &0.01 &0.038& -& 2.0\%\\
\textrm{\cite{balzer2016numerical} - Numerical}& 189 & 0.0089 & -& -& -\\
present& 198 & 0.0095 &0.033& 0.036& 16.9\%\\
\end{tabular}
\end{table}

\section{\label{sec:SPOD} Dominant coherent structures in the LSB}
The role of coherent structures in the dominant dynamics of the LSB flow can be studied through a SPOD analysis. The SPOD provides a means of identifying spatially and temporally correlated structures in a statistically stationary flow. To perform the SPOD, we create a database of approximately $ 7{,}000$ snapshots containing the three components of the 3D velocity field covering a convective time span of $tU_{\infty}/\delta_1^*\approx 3645$. The instantaneous 3D velocity fields are decomposed into a set of spatial Fourier modes along the homogeneous spanwise direction using the discrete Fourier transform as
\begin{eqnarray}
    \boldsymbol{u}(x,y,z,t) = \sum_m \hat{\boldsymbol{u}}_m(x,y,t)\, \mathrm{e}^{\mathrm{i}m\beta_0 z}.
\end{eqnarray}
Here, $m$ is an integer, and $\beta_0 = 0.036\pi/\delta_1^*$ is the fundamental wavenumber set by the spanwise extent of the computational domain. This yields 2D Fourier modes $\hat{\boldsymbol{u}}_m(x,y,t)$ in the $x-y$ plane at each sampled time step. We restrict the analysis to $m\leq 6$, as the modal energy decreases progressively with wavenumber and the large-scale coherent structures of interest reside at these lower wavenumbers; higher wavenumbers are also less reliably resolved on the present grid. For each $m$, we then divide the total snapshots of $\hat{\boldsymbol{u}}_m(x,y,t)$ into $N_b=3$ and $12$ blocks with $50\%$ overlap, each containing $N_s=3500$ and $1024$ snapshots, respectively. We find that the trends in energy and mode shapes do not differ significantly between the $3$- and $12$-block decompositions. However, we report both eigenspectra, since low-frequency dynamics ($St<0.002$) are analyzed using results from $N_b=3$, whereas better-converged results from $N_b=12$ are used for $St>0.002$.

\begin{figure}
\centering
\includegraphics[width=0.95\textwidth]{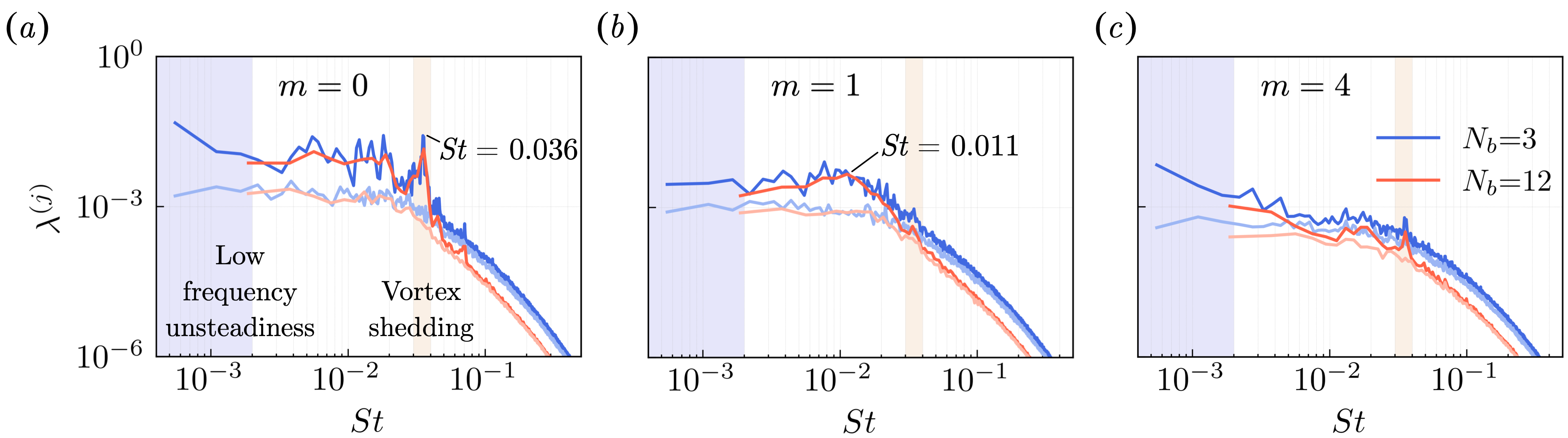}
\caption{\label{fig:6} The SPOD eigenspectra of the LSB flow at representative spanwise wavenumbers $m\beta_0$ with (a) $m=0$, (b) $m=1$, (c) $m=4$. The eigenvalues are shown in progressively lighter shades of blue and orange from the first ($j=1$) towards the second ($j=2$) mode.}
\end{figure}

The SPOD eigenspectra at three representative spanwise wavenumbers $m\beta_0$ (harmonics $m=0,1,4$) are shown in Fig.~\ref{fig:6}. Here, the eigenvalues ($\lambda^j$) represent the spectral density of the kinetic energy of the corresponding SPOD mode associated with a particular wavenumber and frequency.
At $m=0$ (Fig.~\ref{fig:6}a), the leading eigenvalue $\lambda^1$ exhibits a dominant peak at the vortex shedding frequency $St=0.036$ and remains well separated from $\lambda^2$ across the range $St<0.02$. This large separation ($\lambda^1\gg\lambda^2$) indicates rank-one behavior \citep{towne2018spectral}, so that the first SPOD mode captures the dominant structure at these frequencies. Similar to the PSD of the probe data, $\lambda^1$ (obtained with $N_b=3$) also rises for frequencies below $St=0.002$ while retaining the rank-one feature, indicating that the low-frequency unsteadiness is associated with coherent structures rather than broadband turbulence. In contrast, at $m=1$ (Fig.~\ref{fig:6}b), the dominant energy ($\lambda^1$) concentrates in a localized frequency range centered at $St=0.011$, with no conspicuous elevation at the shedding frequency ($St=0.036$) or in the low-frequency range ($St<0.002$), indicating the absence of those mechanisms at this wavenumber. At $m=4$ (Fig.~\ref{fig:6}c), spikes in $\lambda^1$ reappear at the shedding frequency and across $St<0.002$, although at lower levels than those at $m=0$, signaling the presence of coherent 3D structures at these frequencies. We note, however, that both vortex shedding and low-frequency unsteadiness retain significant energy in the range $3\leq m\leq 5$ (see Fig.~\ref{fig:appB}), so these dynamics are distributed across a range of spanwise wavenumbers.

\begin{figure}
\centering
\includegraphics[width=0.95\textwidth]{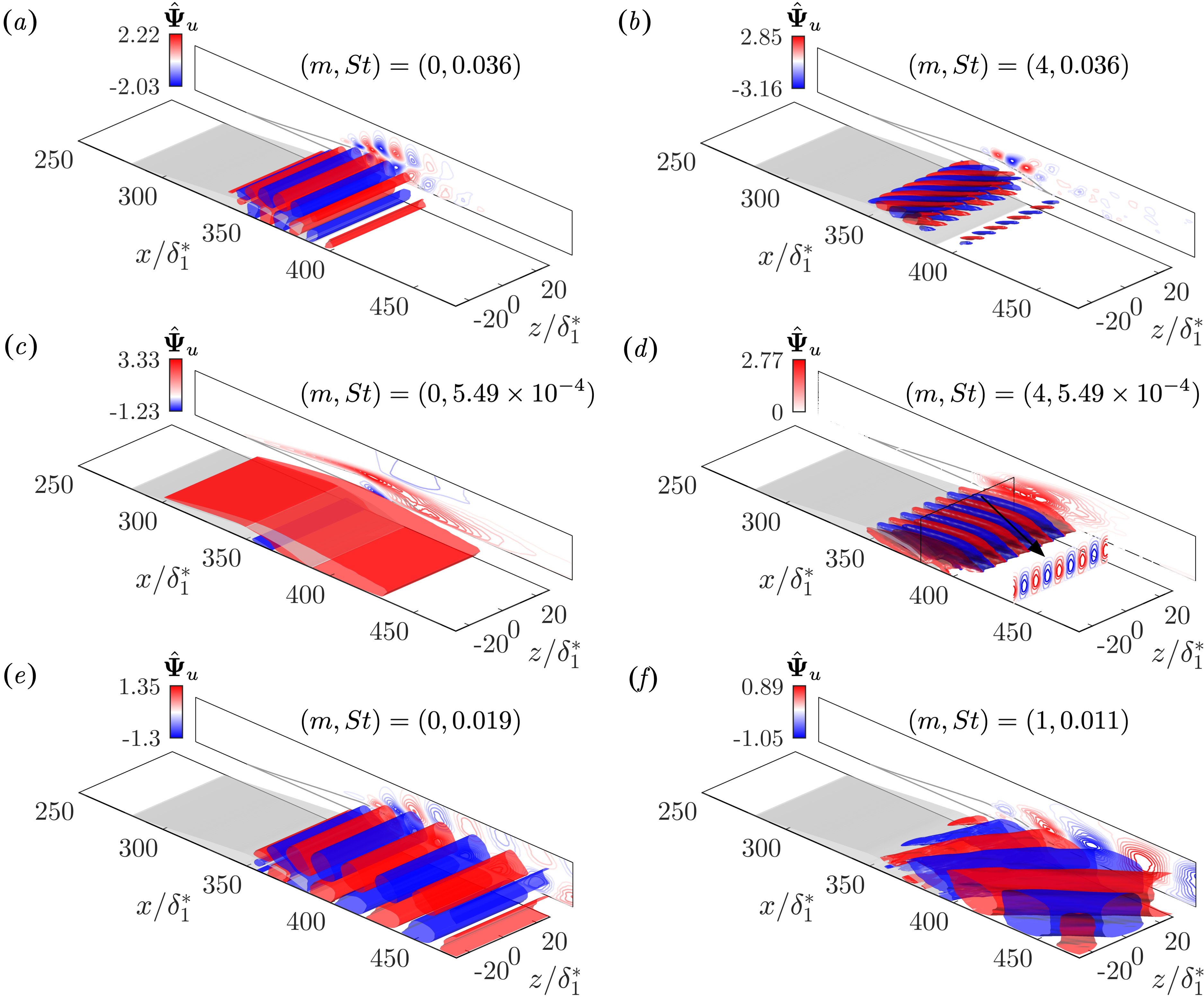}
\caption{\label{fig:7} The streamwise velocity component of the leading SPOD mode at pairs of wavenumber harmonic and frequency $(m, St)$ of (a) $(0,0.036)$, (b) $(4,0.036)$, (c) $(0,5.49\times 10^{-4})$, (d) $(4,5.49\times 10^{-4})$, (e) $(0,0.019)$, and (f) $(1,0.011)$.}
\end{figure}

It is now instructive to examine the SPOD modes to identify spatially and temporally coherent regions in the flow associated with the dominant frequencies and wavenumbers identified in the eigenspectra (Fig.~\ref{fig:6}). The streamwise velocity component ($\hat{u}'_{\text{SPOD}}$) of the first SPOD mode is shown in Fig.~\ref{fig:7} at selected frequencies and wavenumber harmonics. The SPOD mode at the vortex shedding frequency ($St=0.036$) in Fig.~\ref{fig:7}(a) shows the presence of KH-type traveling-wave structures along the separated shear layer for $m=0$. Saturation of these nominally 2D KH waves produces shear-layer roll-up and downstream vortex shedding. Since no upstream forcing is introduced in the LES, the self-excited vortex shedding appears likely due to a region of absolute instability of the KH wave within the LSB caused by strong reverse flow. The representative 3D ($m=4$) SPOD mode at the vortex shedding frequency (Fig.~\ref{fig:7}(b)) shows coherent structures in the form of oblique ($m\neq 0, \, St\neq 0$) KH waves that are present in the region where the modulation of the 2D vortex roller occurs. Although shown here only at the wavenumber corresponding to $m=4$, similar oblique-wave structures are present across the range of wavenumbers for $3\leq m\leq5$ at the frequency $St=0.036$. Such oblique-wave disturbances have been reported to play an important role in vortex breakdown and transition in the LSB \citep{michelis2018origin}. Moreover, the selection of the dominant wavenumber of the amplified oblique wave is non-deterministic and depends on the background disturbance \citep{michelis2018origin}. 

The SPOD spectrum in the low-frequency unsteadiness range ($St<0.002$) does not have any well-defined peak. Therefore, we show the SPOD mode at the lowest resolved frequency, $St=5.49\times 10^{-4}$, as representative of that range. However, the true dominant frequency of these dynamics may well be lower than this resolved limit. The low-frequency SPOD mode at $m=0$ in Fig.~\ref{fig:7}(c) is associated with coherent velocity fluctuations along the separated shear layer, reflecting a low-frequency streamwise oscillation of the bubble. The fluctuations near the flat plate are enhanced around the mean reattachment location, suggesting an unsteady excursion of the reattachment point about its mean position. In contrast, the separation location remains nearly constant, as evidenced by the negligible velocity fluctuations. The global low-frequency oscillation of the separated zone is consistent with the flapping behavior found in previous studies of the LSB \citep{zaman1989natural,michelis2017response}. The SPOD mode at the same low frequency but at the spanwise wavenumber with $m=4$ in Fig.~\ref{fig:7}(d) shows coherent structures resembling the streamwise-elongated streaks typically observed in attached boundary-layer flow \citep{ran2019stochastic}. These structures are localized in the aft portion of the LSB, with alternating signs of velocity fluctuations along the spanwise direction, clearly observed in a planar view sliced at $x/\delta_1^*=380.6$. Recent studies of the flat-plate LSB have also reported streaks near reattachment \citep{malmir2024low,borgmann2025experimental}, linking their formation and advection to the bubble's low-frequency expansion and contraction. The origin of these streaks, however, remains unconfirmed. Additionally, we show the SPOD mode at $(m, St)=(0,0.019)$ and $(m, St)=(1,0.011)$ located at intermediate frequencies between the low-frequency unsteadiness and the vortex shedding in Fig.~\ref{fig:7}(e) and (f), respectively. Both modes possess a rank-one feature, with the mode at $(m, St)=(0,0.019)$ possibly linked to a subharmonic KH instability due to the temporal frequency being approximately half of the vortex shedding frequency, and the mode corresponding to $(m, St)=(1,0.011)$ associated with the breakdown of the spanwise rollers downstream of the reattachment zone. We therefore exclude both modes from further discussion in the paper, as the structures are mostly present downstream of the bubble and likely do not play a significant role in the transition process within the LSB.

\section{\label{sec:resolvent} Physics-based modeling of the coherent structures}
The coherent structures identified by SPOD can arise from the linear amplification of disturbances \citep{towne2018spectral}. For this reason, in this section we focus on exploring physics-based modeling of the linear mechanisms underlying the coherent structures using resolvent analyses. First, we model the linear amplification of perturbations around a statistically stationary LSB flow using the CRA, and subsequently employ mean and harmonic resolvent analyses to capture the effect of base-flow unsteadiness on this amplification.

\subsection{Perturbation amplification about stationary base flow: CRA}
We begin our analysis by examining the eigenvalues of the mean-flow-based linear operator $\boldsymbol{A}(\overline{\boldsymbol{Q}},\beta)$ by solving the following
\begin{equation}
    \boldsymbol{A}(\overline{\boldsymbol{Q}},\beta) \boldsymbol{\phi} = \omega_{\text{eig}} \boldsymbol{\phi},
\end{equation}
where $\omega_{\text{eig}} =\omega_{\text{eig},r} + \mathrm{i} \omega_{\text{eig},i}$ is a complex eigenvalue and $\boldsymbol{\phi}$ is the corresponding eigenvector. The reason for examining the eigenvalues is to find the presence of any marginally stable ($\omega_{\text{eig},r}=0$) or unstable ($\omega_{\text{eig},r}>0$) eigenvalues, since the resolvent operator $[s\boldsymbol{I}-\boldsymbol{A}(\overline{\boldsymbol{Q}},\beta)]^{-1}$ is ill-posed if $s = \alpha + \mathrm{i}\omega$ lies near $\omega_{\text{eig}}$. Note that we introduce the complex frequency $s$ in the resolvent operator, which, for $\alpha=0$, reduces to the standard definition of the resolvent operator given earlier. In such a case ($\omega_{\text{eig}}=s$), the analysis yields an infinite gain, although in practice a sharp peak appears in the resolvent gain spectrum, with the eigenvector as the response. To regularize the problem, a discounting approach \citep{jovanovic2004modeling,rolandi2024invitation,sun2020resolvent} can be used as one of the few available methods \citep{colonius2025modal}, by setting $\alpha>\max(\omega_{\text{eig},r})$. 
 
We note that the current base flow used to generate the linearized operator $\boldsymbol{A}(\overline{\boldsymbol{Q}},\beta)$ is a spanwise- and time-averaged flow, which is strictly speaking not a fixed point of the Navier--Stokes equation and contrary to the traditional idea of linear stability analysis. Here, the eigenvalue spectrum will merely reflect the properties of the linear operator $\boldsymbol{A}(\overline{\boldsymbol{Q}},\beta)$ linearized around the current base flow $\overline{\boldsymbol{Q}}$; as such, any eigenvalues with positive growth rates ($\omega_{\text{eig},r}>0$) correspond to global modes that are amplified or that the system would likely amplify, rather than traditional asymptotic stability properties. Nevertheless, we denote the eigenvalues in the following discussions as unstable/stable, but the reader should interpret the result as pertaining to the properties of this specific operator $\boldsymbol{A}(\overline{\boldsymbol{Q}},\beta)$.

\begin{figure}
\centering
\includegraphics[width=0.95\textwidth]{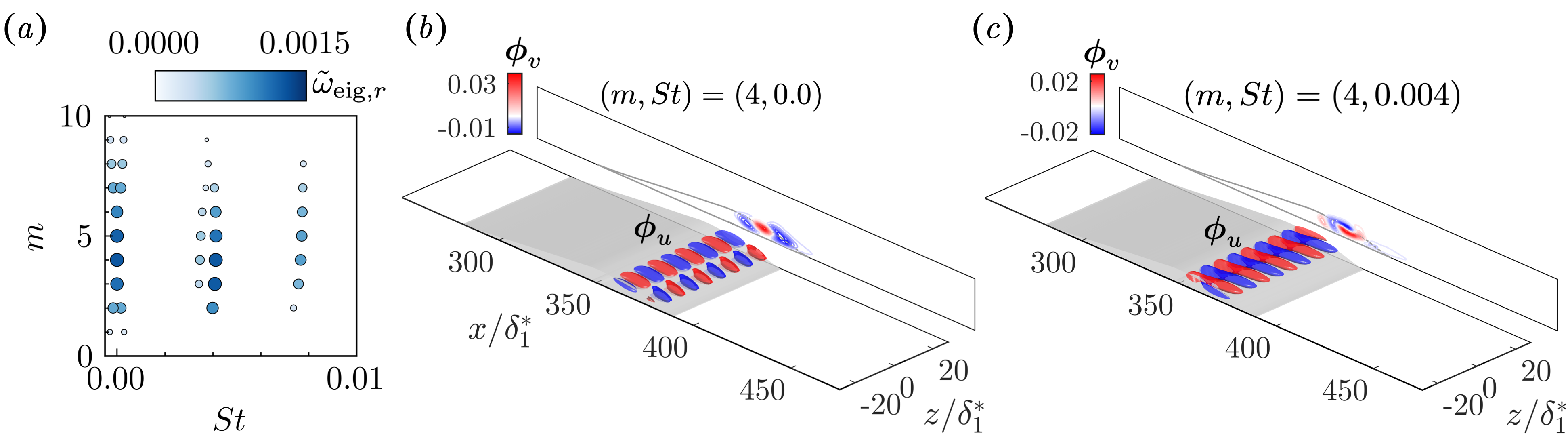}
\caption{\label{fig:8} (a) Eigenvalues with positive real part of the mean flow-based linear operator $\boldsymbol{A}(\overline{\boldsymbol{Q}},\beta)$ across a range of spanwise wavenumbers $\beta = m\beta_0$. Isosurface of the streamwise velocity, and contour of the wall-normal velocity component of the eigenmode at (b) $(m, St)=(4,0.0)$, and (c) $(m, St)=(4,0.004)$.}
\end{figure}

The eigenvalue spectra for spanwise wavenumbers $m\beta_0$, with $\beta_0 \delta_1^*=0.036\pi$, in the range $0 \leq m \leq 10$ are plotted in Fig.~\ref{fig:8}(a). Only eigenvalues with positive growth rate ($\omega_{\text{eig},r}> 0$) are shown. The eigenvalues are colored by their normalized growth rate $\tilde{\omega}_{\text{eig},r} = \omega_{\text{eig},r} \delta_1^*/2\pi U_{\infty}$, the values of which do not carry the same meaning as in the classical stability analysis due to the usage of mean flow as the base. The imaginary part of the eigenvalue ($\omega_{\text{eig}, i}$) corresponds to the temporal frequency and is normalized as the Strouhal number $St=\omega_{\text{eig}, i} \delta_1^*/2\pi U_{\infty}$, which can be related to instabilities present in the nonlinear flow. As evident in Fig.~\ref{fig:8}(a), the linear operator is globally stable for 2D ($m=0$) perturbations. The absence of an unstable or marginally stable global mode for 2D perturbations, despite vortex shedding in the nonlinear flow, has also been reported for a flat-plate LSB \citep{borgmann2025experimental} and for airfoil flow, where \cite{rolandi2025biglobal} attributed it to the use of the mean flow for linearization. 

The linear operator possesses unstable 3D global modes for all the nonzero spanwise wavenumbers in the range $1\leq m \leq 10$ (Fig.~\ref{fig:8}(a)), implying a possibility of the three-dimensionalization of the flow through modal instabilities. The unstable eigenvalues are clustered near three frequency branches at $St\approx0$, $St\approx0.004$, and $St\approx0.0075$, all of which are orders of magnitude lower than the vortex shedding frequency of $0.036$. The maximum eigenvalue growth rate, $\max(\tilde{\omega}_{\text{eig},r})\approx 0.0013$, is attained at the spanwise wavenumber corresponding to $m=4$. We show the isosurface of the real part of the global modes ($\boldsymbol{\phi}_u$) along with contours of $\boldsymbol{\phi}_v$ at this wavenumber ($m=4$) in Fig.~\ref{fig:8}(b-c) at frequencies $St=0$ and $St=0.004$. The spatial distribution of the eigenmode at each frequency is concentrated in the aft portion of the LSB near the concave streamline curvature, indicating a centrifugal mechanism as the likely cause for the existence of these modal instabilities \citep{rodriguez2010structural,cherubini2010onset}. The wall-normal velocity ($\boldsymbol{\phi}_v$) shapes are in good agreement with the most unstable 3D eigenmodes reported by \cite{borgmann2025experimental} using a dynamic mode decomposition-based linear stability analysis. We note that the instability of the stationary ($St=0.0$) global modes can act as the primary mechanism (for $\overline{u}_{\text{rev}}/U_{\infty}<16\%$) to instigate the three-dimensionalization of the flow \citep{rodriguez2021self}. We consider this unlikely in the present case, given that the mean peak reverse flow reaches $\approx 21.4\%$ of the free stream, making the KH instability the primary mechanism. Also, non-modal amplification of perturbation can render the flow three-dimensional before the centrifugal instabilities become active, which we can identify from the resolvent analysis. 

The eigenspectra of $\boldsymbol{A}(\overline{\boldsymbol{Q}},\beta)$ revealed unstable eigenvalues, so we adopt a discounting approach \citep{yeh2020resolvent,sun2020resolvent,jovanovic2004modeling,rolandi2024invitation} for the resolvent analysis by adding an offset $\alpha\neq0$ to the complex frequencies. The selected value of $\tilde{\alpha}=0.0022$ is greater than the $\max{(\tilde{\omega}_{\text{eig},r})}=0.0013$, where $\tilde{\alpha}=\alpha \delta_1^*/2\pi U_{\infty}$, rendering the system as stable. Note that the choice of $\tilde{\alpha}$ is arbitrary as long as it is higher than $\tilde{\omega}_{\text{eig},r}$, and the resolvent gain spectrum will be different based on the chosen $\tilde{\alpha}$. We could have a slightly lower $\tilde{\alpha}$, but we choose it to be consistent with the discounting values applied to other resolvent operators later. In the present case, we find that increasing $\tilde{\alpha}$ reduces the resolvent gain across all frequencies, making any peak in the gain distribution flatter, consistent with observations in other studies \citep{rolandi2025biglobal}. The discounting factor is often interpreted as an effective time window $\tilde{t}_{\alpha} = t U_{\infty}/\delta_1^* = 1/2\pi \tilde{\alpha}$ over which the perturbation gets amplified. Based on the chosen discounting value, this time window corresponds to $\tilde{t}_{\alpha}\approx 72$. With the discounting applied the modified classical resolvent operator reads as $[(\alpha+\mathrm{i}\omega)\boldsymbol{I} - \boldsymbol{A}(\overline{\boldsymbol{Q}},\beta)]^{-1}$, and the SVD of this modified operator is used to study the non-modal perturbation amplification within the time window $\tilde{t}_{\alpha}$.

\begin{figure}
\centering
\includegraphics[width=0.95\textwidth]{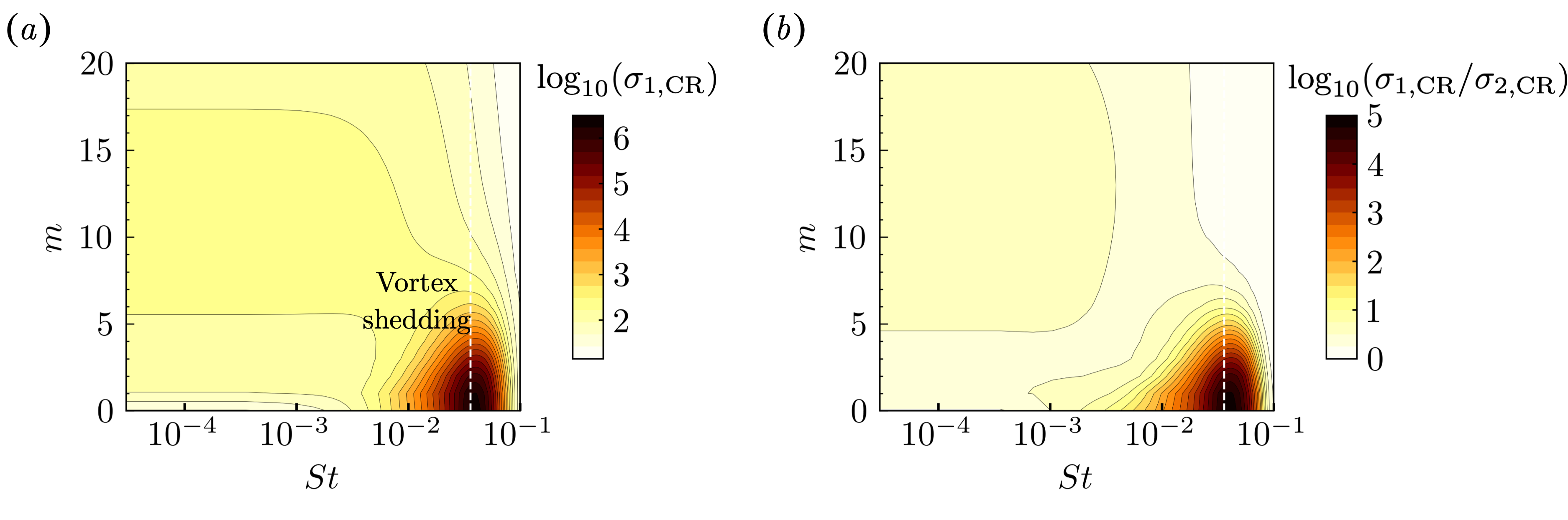}
\caption{\label{fig:9} Map of the (a) optimal singular value ($\sigma_{1,\text{CR}}$) obtained using CRA, and (b) the ratio of the optimal to the first sub-optimal singular value ($\sigma_{1,\text{CR}}/\sigma_{2,\text{CR}}$), over a range of frequency and wavenumber.}
\end{figure}

We perform the CRA for pairs of temporal-frequency and spanwise-wavenumber harmonics over the ranges $0\leq St\leq 0.1$ and $0\leq m\leq 20$, respectively. The contours of the optimal singular value $\sigma_{1,\text{CR}}$ in the wavenumber-frequency ($m$--$St$) space are shown in Fig.~\ref{fig:9}(a). A zone of significant linear amplification with $\sigma_{1,\text{CR}}$ on the order of $10^4-10^6$ is evident within $0.032 \leq St \leq 0.045$ and $0\leq m\leq 5$. Since these frequencies are close to the vortex-shedding frequency at $St=0.036$, the amplification mechanism in this zone is attributed to the KH instability. The highest amplification occurs for the planar ($m=0$) perturbation at $St=0.04$. The singular values decrease progressively with increasing $m$, indicating that the amplification of oblique KH waves is secondary to that of the planar KH wave. Moreover, amplification of oblique KH waves can be seen (Fig.~\ref{fig:9}(a)) to occur across a range of wavenumbers below $m=6$, agreeing with the observation from SPOD. The amplification across a broad range of frequencies and wavenumbers highlights the intricate nature of the LSB flow, in which many structures may be amplified and interact, and their eventual dominance is selected through the receptivity process.

The ratio between the leading two singular values ($\sigma_{1,\text{CR}}/\sigma_{2,\text{CR}}$) additionally reveals the nature of perturbation amplification. If the flow is dominated by a strong convective instability mechanism, the optimal singular value dominates, i.e., the ratio $\sigma_{1,\text{CR}}/\sigma_{2,\text{CR}}$ becomes larger \citep{beneddine2016conditions}. Also, a larger $\sigma_{1,\text{CR}}/\sigma_{2,\text{CR}}$ indicates a rank-one characteristic of amplification, i.e., the optimal forcing and response capture most of the linear amplification dynamics. As the planar and oblique KH waves are linearly amplified convective instabilities, we notice a large separation between the leading two singular values on the order $\sim10^4$ and higher in Fig.~\ref{fig:9}(b) in the same zone in $m$--$St$ space where the dominant amplification occurs (Fig.~\ref{fig:9}(a)). The large separation also indicates a rank-one amplification feature of the KH waves, similar to what we found from the SPOD eigenvalues. 

\begin{figure}
\centering
\includegraphics[width=0.95\textwidth]{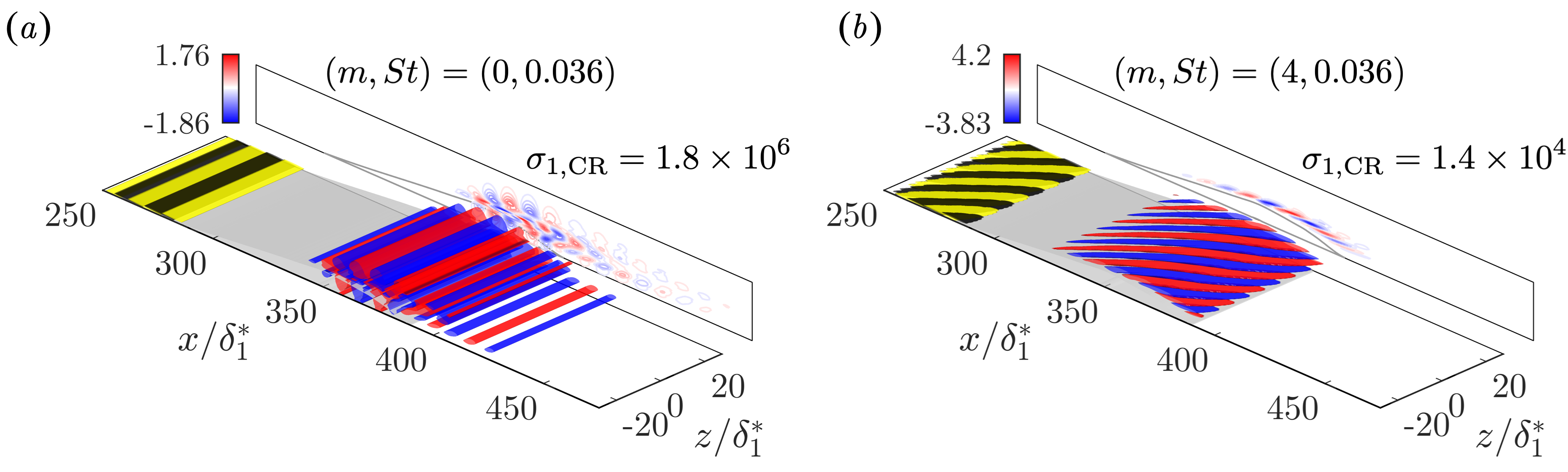}
\caption{\label{fig:10} Isosurface of the real part of the streamwise velocity component of the optimal forcing and response mode at (a) $(m,St)=(0,0.036)$ and (b) $(m,St)=(4,0.036)$, showing the planar and oblique KH wave amplification.}
\end{figure}

We plot in Fig.~\ref{fig:10}(a) the spatial structures of the streamwise velocity component of the optimal forcing and response mode at $(m, St)=(0,0.036)$ and $(m, St)= (4,0.036)$ located within the zone of most amplified perturbations. Both modes show spatial locations of the forcing upstream of separation, with the amplified response mostly along the separated shear layer, and the spatial separation between them is consistent with the convective nature of the amplification. The amplification of these KH waves in the shear layer also agrees with the corresponding SPOD mode at the same wavenumber-frequency pair (see Fig.~\ref{fig:7}(a-b)), although the structures do not appear to be identical. The modes captured by the SPOD in the nonlinear flow are the saturated form, developing from both linear amplifications and nonlinear interactions with other scales, and thus may differ from the optimal resolvent response mode. 

In addition to the dominant KH wave amplification, we also identified a significant presence of fluctuating kinetic energy at frequencies $St<0.002$ in the SPOD eigenspectra (see Fig.~\ref{fig:6}). However, the optimal singular values in Fig.~\ref{fig:9}(a) across all wavenumbers for $St<0.002$ are on the order of $\sim10^2$ or less, suggesting weak non-modal amplification of perturbations. The leading two singular values are also not well-separated, as evident from Fig.~\ref{fig:9}(b), implying the absence of a dominant rank-one linear amplification mechanism in this frequency range ($St<0.002$). In light of the apparent discrepancy between the findings from SPOD and resolvent in the low-frequency zone, we recall that the flow response can be reconstructed from the singular value decomposition of the resolvent operator as
\begin{eqnarray}
    \hat{\boldsymbol{q}}' = \sigma_1 \boldsymbol{\psi}_1 \langle\boldsymbol{\varphi}_1,\hat{\boldsymbol{f}}'\rangle + \sum_{i\geq2} \sigma_i \boldsymbol{\psi}_i \langle\boldsymbol{\varphi}_i,\hat{\boldsymbol{f}}'\rangle,
    \label{eqn:21}
\end{eqnarray}
where $\boldsymbol{\varphi}_i$ and $\boldsymbol{\psi}_i$ are the forcing and response modes ranked by singular values $\sigma_i$. Since no external forcing is introduced in the present simulation, $\hat{\boldsymbol{f}}'$ represents the intrinsic nonlinear forcing generated by the flow itself. We highlight that the energy in the response depends on two distinct factors: the gain $\sigma_i^2$ and the projection of the intrinsic forcing onto the forcing mode $\langle\boldsymbol{\varphi}_i,\hat{\boldsymbol{f}}'\rangle$.
We can combine these two factors into a coefficient \citep{lugrin2021transition} that quantifies the energy amplified through the optimal mode alone as
\begin{eqnarray}
    c = \sigma_1^2 |\langle\boldsymbol{\varphi}_1,\hat{\boldsymbol{f}}'\rangle|^2,
\end{eqnarray}
where $\sigma_1^2$ captures the ability of the resolvent operator to optimally amplify the structures through a linear mechanism at a specific wavenumber-frequency pair, whereas $|\langle\boldsymbol{\varphi}_1,\hat{\boldsymbol{f}}'\rangle|^2$ captures the strength of the excitation, i.e., how effective the nonlinear intrinsic forcing is in driving the amplification of those structures. We note that the projection term $|\langle\boldsymbol{\varphi}_1,\hat{\boldsymbol{f}}'\rangle|^2$ is understood in a statistical sense, averaged over realizations. If $\sigma_1$ is large, then the value of $c$ will also be large even if $\langle\boldsymbol{\varphi}_1,\hat{\boldsymbol{f}}'\rangle$ is small. This is the case for the KH wave amplification zone in the $m$--$St$ space, where both $c$ and SPOD energy are high. The rank-one nature of the response leads to dominance by the optimal mode, and any intrinsic forcing with even a modest component along $\boldsymbol{\varphi}_1$ will be amplified by a factor of $\sigma_1^2\approx10^8-10^{12}$. In this regime, the resolvent and the SPOD modes agree well because the amplification gain is sufficiently large to optimally excite the structures.

In the case of weak amplification, i.e., small $\sigma_1^2$, but high SPOD energy as for $St<0.002$, two possibilities exist: In scenario (i), if no dominant linear amplification mechanism is present, then $c$ is not expected to match the SPOD energy, since there is no a priori reason for the intrinsic forcing to align with the optimal forcing direction $\boldsymbol{\varphi}_1$. The nonlinear interactions must then play an important role in exciting coherent structures that lead to the high SPOD energy. In scenario (ii), if the forcing instead projects strongly onto $\boldsymbol{\varphi}_1$, then $c$ can attain a large value, and the observed high SPOD energy stems from a weak, but strongly excited, linear mechanism. 

\begin{figure}
\centering
\includegraphics[width=0.95\textwidth]{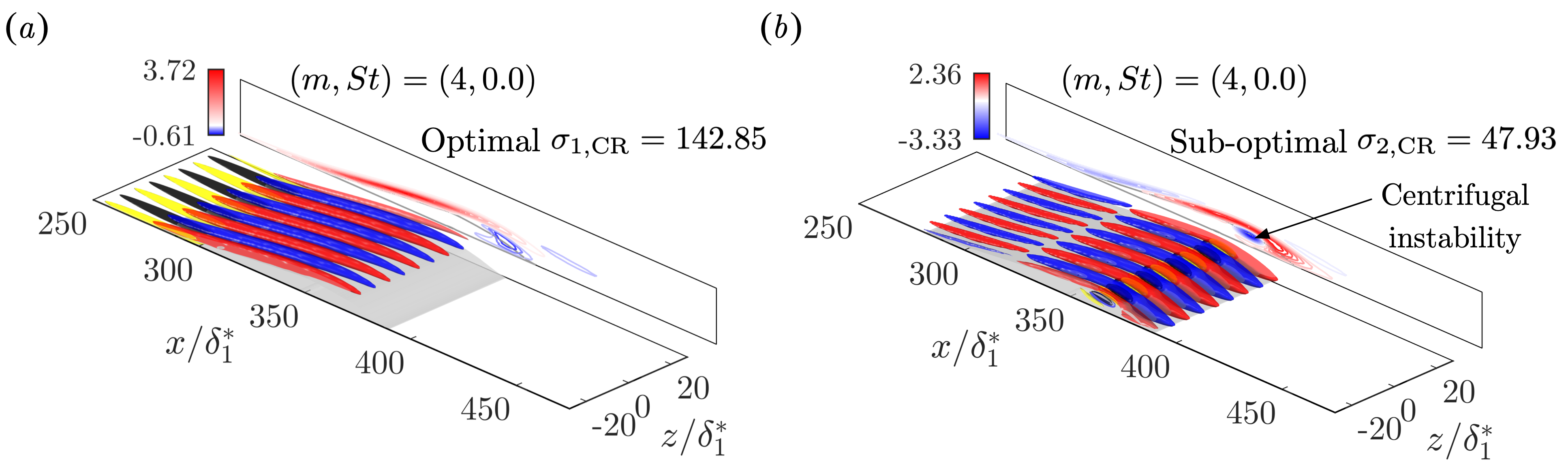}
\caption{\label{fig:11} Isosurface of the real part of the streamwise velocity component of the (a) optimal and (b) sub-optimal forcing and response mode at $(m,St)=(4,0.0)$.}
\end{figure}

We argue that no dominant linear amplification mechanism is present in the low-frequency region. To support this claim, we show both the optimal and sub-optimal forcing and response modes at $(m, St)=(4,0.0)$ in Fig.~\ref{fig:11}(a-b). The optimal response structures are streaks in the initial part of the separated shear layer, whereas in the SPOD mode streaks are located near the reattachment zone. This mismatch in the streak location between the optimal response and the SPOD mode indicates that the intrinsic forcing does not simply project onto $\boldsymbol{\varphi}_1$, consistent with scenario (i). Furthermore, the sub-optimal response shows amplification of both streaks along the shear layer and centrifugal instability within the recirculation zone. The absence of a dominant, rank-one linear mechanism therefore favors scenario (i): the streaky structures captured by the SPOD originate in the intrinsic nonlinear forcing rather than in same-frequency linear amplification. In the next section, we show that this forcing is not arbitrary but arises from cross-frequency transfer mediated by base-flow unsteadiness, a linear mechanism inaccessible to the classical resolvent framework.

\subsection{Perturbation amplification about unsteady base flow: MRA and HRA}
The SPOD and CRA identified the 2D KH instability as the dominant mechanism leading to vortex shedding in the nonlinear flow. In this section, we model the perturbation dynamics induced by interactions with both the mean flow and an unsteady component, which we represent using the Fourier mode at the vortex-shedding frequency. The normalized Fourier coefficients of the state variables $[\rho,\rho u,\rho v,\rho w,\rho E]$ at the vortex-shedding frequency $\omega_0=St_0(2\pi U_{\infty})/\delta_1^*$, where $St_0=0.036$, are computed from the LES snapshots following
\begin{equation}
    \hat{\boldsymbol Q}(\omega_0) = \frac{1}{N}\sum_{k=0}^{N-1} 
    \boldsymbol{Q}(k)\, \mathrm{e}^{-\mathrm{i}\omega_0 k\,\Delta t},
\end{equation}
where $\Delta t$ is the sampling time interval of the snapshots. Here, we used $N=6000$ snapshots that span approximately $110$ cycles of vortex shedding. The time-periodic base flow is then modeled as $\boldsymbol{Q}(t)=\overline{\boldsymbol{Q}}+ \hat{\boldsymbol{Q}}(\omega_0)\,\mathrm{e}^{\mathrm{i}\omega_0 t} +\,\mathrm{c.c.}$, leading to the set of base-flow frequencies $\Omega_b=\{-\omega_0,0,\omega_0\}$ for linearization of the dynamics. Using 
this base flow, we generated the linear operators $\hat{\boldsymbol{A}}_0(\beta)$ and $\hat{\boldsymbol{A}}_{\pm\omega_0}(\beta)$ in the frequency domain and assembled the operator $\boldsymbol{T}(\beta)$ (see Eq.~(\ref{optT})). We considered the set of frequencies $\Omega_{\gamma}=\gamma\omega_0+\{-5,\dots,-1,0,1,\dots,5\}\omega_0$ to represent the perturbations in the frequency domain.

As a precursor to the resolvent analysis, we examine the eigenvalues of the linear operator $\boldsymbol{T}(\beta)$ to identify any unstable modes. We solve the eigenvalue problem obtained by setting the forcing 
$\hat{\boldsymbol{\mathcal{F}}}'_{\beta,\gamma}$ in Eq.~(\ref{eqn10}) to zero. The eigenvalues are computed for spanwise wavenumber harmonics in the range $0\leq m\leq5$, since both SPOD and CRA showed these wavenumbers to be important. The eigenvalues revealed several unstable modes, which we do not show here for brevity, with the highest normalized growth rate (positive real part) of $0.0018$. Due to the unstable eigenvalues, we adopt a discounting method, similar to the CRA, with discount parameter $\alpha=0.0022$. The chosen value exceeds the maximum eigenvalue growth rate, constraining the perturbation 
amplification over the same time window as in the CRA, and thereby allowing a direct comparison between the amplification about the stationary and time-periodic base flows. The HRA and MRA are performed by computing the SVD of the modified operators $\boldsymbol{H}(\beta,\gamma,\alpha) =[(\alpha+\mathrm{i}\gamma\omega_0)\boldsymbol{I}-\boldsymbol{T}(\beta)]^{-1}$ and $\boldsymbol{H_{0,0}}(\beta,\gamma,\alpha) = \boldsymbol{P}_y^{T}\boldsymbol{H}(\beta,\gamma,\alpha)\boldsymbol{P}_f$, respectively. We note that after constructing the operator $[(\alpha+\mathrm{i}\gamma\omega_0)\boldsymbol{I}-\boldsymbol{T}(\beta)]$, an LU decomposition needs to be performed only once and can then be reused to solve the linear systems within the randomized SVD algorithm \citep{ribeiro2020randomized} for both resolvent operators, keeping the computational cost feasible.

\begin{figure}
\centering
\includegraphics[width=0.95\textwidth]{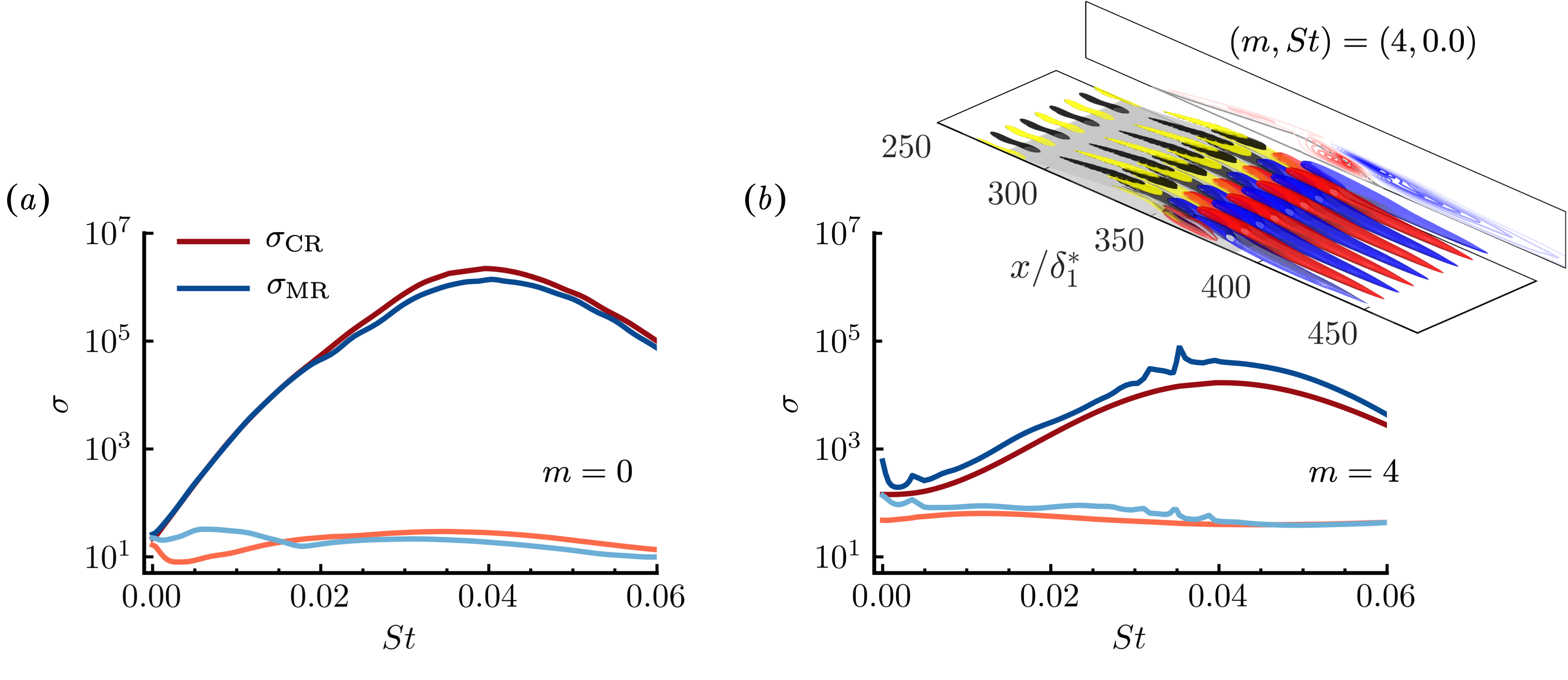}
\caption{\label{fig:12} The distribution of the leading two singular values obtained from the CRA and MRA over a range of frequencies at spanwise wavenumber harmonics (a) $m=0$ and (b) $m=4$. The isosurface of the optimal MR forcing and response at $(m, St)=(4,0.0)$ is also shown in (b) using the real part of the streamwise velocity component.}
\end{figure}

We begin the discussion of perturbation amplification by the unsteady base flow $\boldsymbol{Q}(t)$ by examining the leading two singular values for amplification at the same frequency in the input and output, obtained from the MRA in Fig.~\ref{fig:12}(a--b). Because the amplification gain ($\sigma^2_{1,\text{MR}}$) relates to single-frequency input-output, these singular values can be compared directly with those from the CRA. We recall that the MRA amplification arises from an unsteady base flow, whereas in the CRA the base flow is stationary. Despite this difference, the optimal singular value distributions at $m=0$ and $m=4$ are comparable over the considered frequency range: the KH band around $St=0.04$ remains the dominant mechanism, with a rank-one feature evident from the well-separated singular values, while in the low-frequency region ($St<0.02$) the gain is almost two orders of magnitude lower. To this extent, the two base flows yield a similar picture.

The MRA response, however, differs from the CRA in a physically significant way. At $(m, St)=(4,0.0)$ we observe a local peak whose optimal forcing and response (Fig.~\ref{fig:12}(b)) are distinct from the CRA optimal modes (Fig.~\ref{fig:11}(a)). The response streaks, amplified by the unsteady base flow, are located near reattachment and further downstream, and the optimal forcing now resides mostly within the separated shear layer, in addition to upstream of separation. Although not identical to the SPOD mode, the streak location near reattachment and the shift of the optimal forcing into the shear layer bring the amplified structure into much closer agreement with the SPOD observation than the CRA does. The unsteady base flow therefore reshapes \textit{which} structure is preferentially amplified toward the observed streak, even though the same-frequency gain alone remains too weak to account for the high SPOD energy at this wavenumber and frequency. This points to an amplification route beyond the same-frequency mechanism: the streak must be energized via cross-frequency interactions mediated by base-flow unsteadiness, which we examine next using the off-diagonal blocks of the harmonic resolvent operator.

\begin{figure}
\centering
\includegraphics[width=0.95\textwidth]{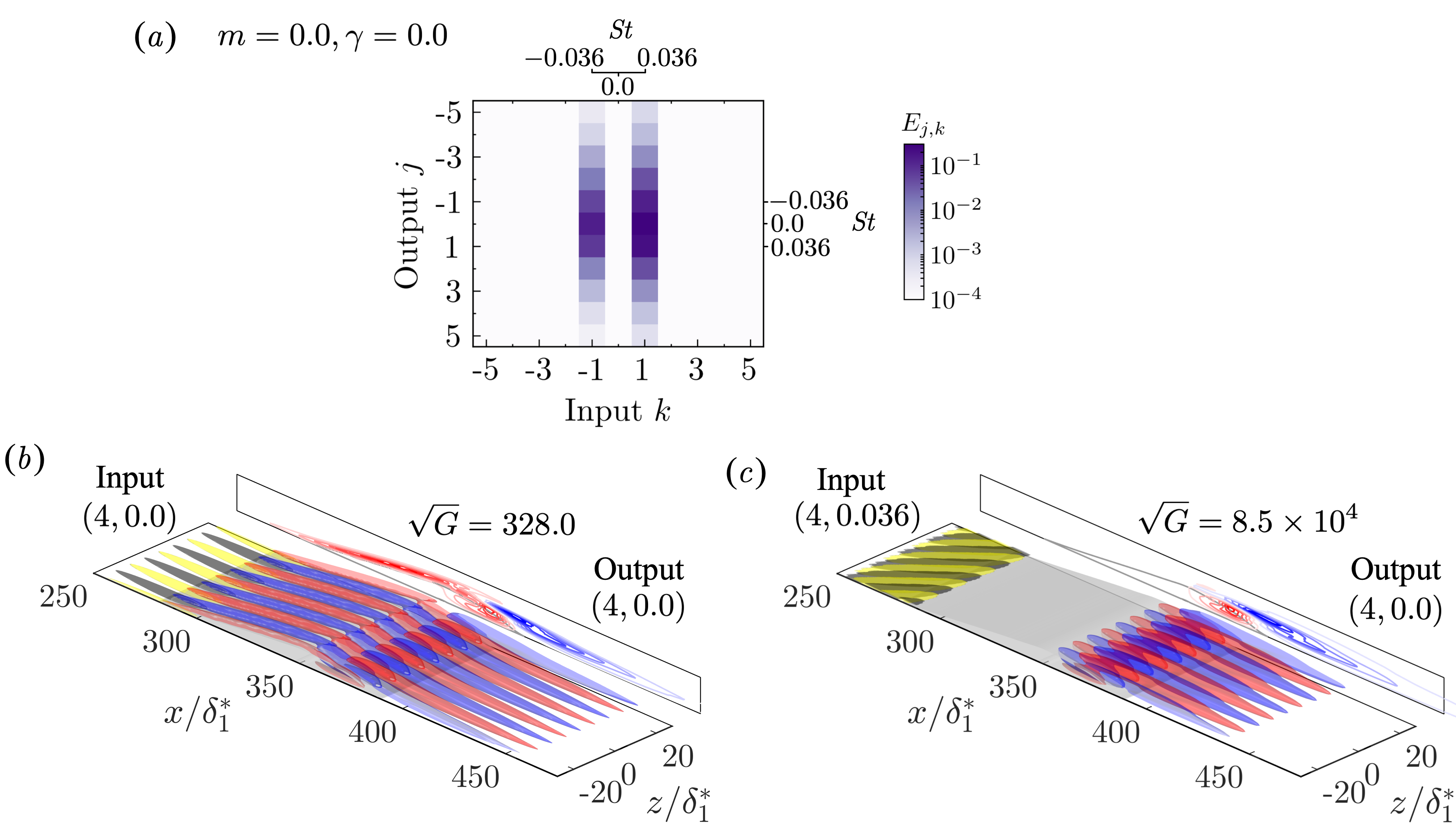}
\caption{\label{fig:13} (a) The amplification through different blocks of the harmonic resolvent operator for spanwise wavenumber with $m=4$ and $\gamma=0$ expressed using the quantity $E_{j,k}$. (b,c) The constrained response at frequency $St=0.0$ to optimal CRA forcing at a single frequency $0.0$ and $0.036$ at the spanwise wavenumber harmonic $m=4$.}
\end{figure}

In the CRA and MRA, we examined perturbation amplification at the same temporal frequency in the input and output by the stationary and unsteady base flow. An unsteady base flow, however, generates a response at multiple frequencies in response to forcing at a single frequency. To quantify how much the input and output at different pairs of temporal frequencies within a set $\Omega_{\gamma}$ are amplified through direct and cross-frequency interactions, we define the metric
\begin{eqnarray}
    E_{j,k} = \frac{\sum_{i=1}^5 \sigma_{i}^{2}(\boldsymbol{H}_{j,k}(\beta,\gamma,\alpha))}
                   {\sum_{i=1}^5 \sigma_{i}^{2}(\boldsymbol{H}(\beta,\gamma,\alpha))},
\end{eqnarray}
where $\boldsymbol{H}_{j,k}(\beta,\gamma,\alpha)$ is the sub-block of $\boldsymbol{H}(\beta,\gamma,\alpha)$ that amplifies the input at frequency $(\gamma+k)St_0$ to the output at frequency $(\gamma+j)St_0$. The amplification map given by $E_{j,k}$ is shown in Fig.~\ref{fig:13}(a) for the spanwise wavenumber $m=4$ and the frequency set $\Omega_0=\{-5,\dots,-1,0,1,\dots,5\}St_0$, where a darker color indicates larger amplification through the block. Comparing the columns for $k=0$ and $k=1$, the most effective input frequency in the set is the vortex-shedding frequency ($St=0.036$), rather than the steady ($St=0$) forcing. Notably, this shedding-frequency forcing amplifies the \textit{steady} ($St=0$) response more strongly than the response at its own frequency ($St=0.036$).

This cross-frequency amplification directly resolves the open question from the CRA. To this end, we take the optimal CRA forcings at $(m,St)=(4,0.036)$ and $(4,0.0)$ and construct two unit-norm augmented forcings ($||\hat{\boldsymbol{\mathcal{F}}'}||_E=1$) by placing each at its corresponding frequency index within the set $\Omega_{\gamma}$ for $\gamma=0.0$. We then obtain the response at $St=0.0$ by multiplying each forcing by $\boldsymbol{H}(\beta,\gamma,\alpha)$, and define the square root of the amplification gain $\sqrt{G}$ as the energy norm of the response, which is directly comparable to the optimal singular value of the resolvent operators. As shown in Fig.~\ref{fig:13}(b,c), applying the unit-norm optimal CRA forcing at $(m,St)=(4,0.036)$ to the harmonic resolvent operator produces a steady response at $(4,0.0)$ with $\sqrt{G}\sim10^4$, whereas the same-frequency forcing at $(4,0.0)$ yields only $\sqrt{G}\approx 328$. The stationary streak is therefore amplified far more effectively by cross-frequency transfer from the oblique KH wave than by any same-frequency linear amplification mechanism, a route mediated by the unsteady base flow that the classical resolvent cannot represent.

The mechanism by which the response at $(m,St)=(4,0.0)$ is generated can be summarized as follows. A forcing at $(4,0.036)$ introduced upstream of separation is amplified at the same wavenumber and frequency into an oblique KH wave through the convective mechanism of the mean flow at $(0,0)$. This oblique wave then interacts with the unsteady part of the base flow, the 2D KH wave at $(0,-0.036)$, to generate an intrinsic forcing at $(4,0.0)$, near the maximum bubble height where both waves are active. The mean flow subsequently amplifies this intrinsic forcing, producing the stationary streak response at $(4,0.0)$.

\section{\label{sec:conclusion}Conclusions}
We performed biglobal resolvent analyses to study the linear amplification of perturbations across a range of spanwise wavenumbers and temporal frequencies in an APG-induced LSB over a flat plate, identifying the dominant 2D and 3D mechanisms. The nonlinear flow obtained from the LES exhibits self-sustained vortex shedding at a well-defined frequency in the aft portion of the bubble, whose breakdown triggers transition to turbulence and subsequent reattachment. A low-frequency flapping is also present, evidenced by significant spectral energy at frequencies orders of magnitude below the shedding frequency. Using SPOD on the velocity and wall-pressure data, we identified the dominant sources of unsteadiness as (i) two-dimensional KH traveling waves at the shedding frequency, (ii) three-dimensional oblique waves coinciding with the spanwise modulation of the vortex roller, and (iii) streamwise-elongated streaks near reattachment, associated with the low-frequency flapping.

Classical resolvent analysis of the mean flow shows that any 2D or weakly 3D small-amplitude forcing within a band around the shedding frequency ($0.03<St<0.05$) is strongly amplified by the convective KH mechanism, with optimal gain on the order of $10^8-10^{12}$.  In contrast, the amplification of both 2D and 3D perturbations at low frequency ($St<0.002$) is several orders of magnitude weaker, and no dominant rank-one mechanism is present. This mismatch between the weak low-frequency gain and the significant SPOD energy of the streaks indicates that the streaks are not sustained by same-frequency linear amplification, but are instead energized by the intrinsic nonlinear forcing.

Incorporating the LES Fourier mode at the shedding frequency as the unsteady base flow, we performed mean and harmonic resolvent analyses to determine how this unsteadiness modifies the perturbation dynamics. The mean resolvent gain remains qualitatively similar to the classical result across most of the wavenumber–frequency space, confirming that same-frequency amplification, even when accounting for the unsteady base flow, does not explain the low-frequency streak energy. The harmonic resolvent, however, resolves the missing mechanism: the unsteady base flow associated with the 2D KH wave couples the oblique KH wave at the shedding frequency to a stationary, spanwise-periodic response, generating a streak that is more energetic than the same-frequency response to the stationary 3D forcing. This cross-frequency transfer, inaccessible within the classical framework, identifies base-flow unsteadiness as the mechanism by which shedding energizes the stationary three-dimensional structures observed near reattachment.

These findings highlight the critical role of base-flow unsteadiness in mediating cross-frequency energy transfer and in sustaining three-dimensional structures in the LSB, with direct implications for physics-informed flow control targeting the separated shear layer. Based on our results, obtained solely from input--output optimization in the frequency domain using the different resolvent variants, we draw the following conclusions, some of which are consistent with earlier direct numerical simulation studies of the externally forced LSB \citep{marxen2004effect,rist2006control}:
\begin{enumerate}
    \item Low-frequency ($St<0.005$) 3D forcing introduced upstream of the LSB experiences minimal gain and thus has a negligible effect on the bubble, in contrast to high-frequency ($0.03<St<0.05$) forcing, which is amplified by several orders of magnitude. If, however, the amplitude of the 3D forcing in the attached boundary layer upstream of separation is large enough, it may instead trigger a bypass transition through the transient growth of streaks in the initial part of the separated shear layer.
    \item This also implies that spanwise-periodic steady 3D disturbances, such as surface modifications upstream of separation, are ineffective in modifying the LSB. Steady forcing cannot, by itself, generate the unsteady fluctuations required to trigger transition within the bubble, unless its initial amplitude is sufficiently high to cause a bypass transition. A small-amplitude steady forcing $(\beta,0)$ can, at best, accelerate the growth of an oblique KH wave $(\beta,0.036)$ through its interaction with the 2D KH wave $(0,0.036)$.
    \item Unsteady 3D forcing in the form of oblique KH waves near the vortex-shedding frequency and spanwise wavenumber $m\beta_0$ with $m<6$, introduced upstream of separation with small amplitude, is the most efficient means of modifying the LSB. Through cross-frequency interactions mediated by the unsteady base flow, this forcing generates steady perturbations that reinforce the stationary streak response and may hasten the onset of transition.
\end{enumerate}

\backsection[Acknowledgments]{This material is based upon work supported by the Air Force Office of Scientific Research under award number FA9550-24-1-0136 (Program Officer: Dr.~Gregg Abate). This work used the Anvil supercomputer at Purdue University through allocation PHY250306 from the ACCESS program, which is supported by the U.S. National Science Foundation. We also acknowledge the computing resources provided by Syracuse University Research Computing.}

\appendix

\section{Inviscid solution for 2D flow with blowing-suction at the top}\label{appA}
In this section, we describe the calculations used to obtain the wall-normal velocity profile specified at the top boundary of the domain, which generates the pressure gradient along the flat plate location. We assume uniform, incompressible, and irrotational flow in a 2D rectangular domain. The domain lengths are normalized using the inflow displacement thickness $\delta_1^*$ of the LES. For such a flow, we can define a velocity potential $\Phi$ such that $u=\partial \Phi/\partial x$ and $v=\partial\Phi/\partial y$, which satisfies Laplace's equation
\begin{eqnarray}
\label{eqn:A1}
    \nabla^2\Phi = 0.
\end{eqnarray}
We solve Eq.~(\ref{eqn:A1}) to obtain $\Phi$ subject to the boundary conditions
\begin{eqnarray}
    \Phi \rightarrow x \ \text{as}\ x\rightarrow\pm\infty,\quad
    \frac{\partial \Phi}{\partial y}(x,0)=0,\quad
    \frac{\partial^2 \Phi}{\partial y^2}(x,L_y) = f(x),
\end{eqnarray}
where the function $f(x)$ is expressed as
\begin{eqnarray}
    f(x) = a_1 \exp\!\left(-a_3(x-a_2)^2\right)
         + b_1 \exp\!\left(-b_3(x-b_2)^2\right)
         + c_1 \exp\!\left(-c_3(x-c_2)^2\right),
\end{eqnarray}
with the coefficient values $a_1=-1.41$, $a_2=241.7$, $a_3=0.005$, $b_1=2.03$, $b_2=277.2$, $b_3=0.0165$, $c_1=0.08$, $c_2=344.4$, $c_3=0.00065$. The coefficient values are determined iteratively so that the inviscid pressure obtained from the calculation matches a target pressure distribution. The analytical form of the velocity potential $\Phi$ can be obtained by considering a Fourier transform along the streamwise direction, leading to
\begin{eqnarray}
    \Phi(x,y) = x + 2\,\Re\!\left[\sum_{n=1}^{\infty}
      \frac{\hat{f}_n}{L_x(nk_0)^2}\,
      \frac{\cosh(nk_0 y)}{\cosh(nk_0 L_y)}\,
      \exp(\mathrm{i}nk_0 x)\right],
\end{eqnarray}
where $k_0=2\pi/L_x$ is the fundamental streamwise wavenumber, $L_x\approx14000$ is the streamwise extent of the domain over which the Fourier coefficients are evaluated, and $\hat{f}_n$ are the Fourier coefficients of $f(x)$ given by
\begin{eqnarray}
    \label{eqnA5}
    \hat{f}_n &=& a_1\sqrt{\frac{\pi}{a_3}}\,
        \exp\!\left(\frac{-n^2k_0^2}{4 a_3}\right)\exp(-\mathrm{i}nk_0a_2)
      + b_1\sqrt{\frac{\pi}{b_3}}\,
        \exp\!\left(\frac{-n^2k_0^2}{4 b_3}\right)\exp(-\mathrm{i}nk_0 b_2) \nonumber\\
    & & + c_1\sqrt{\frac{\pi}{c_3}}\,
        \exp\!\left(\frac{-n^2k_0^2}{4 c_3}\right)\exp(-\mathrm{i}nk_0 c_2).
\end{eqnarray}
The streamwise ($u$) and wall-normal ($v$) velocity at any spatial point in the domain can then be obtained as
\begin{eqnarray}
    u(x,y) &=& 1 + 2\,\Re\!\left[\sum_{n=1}^{\infty}
      \frac{\mathrm{i}\hat{f}_n}{nk_0 L_x}\,
      \frac{\cosh(nk_0 y)}{\cosh(nk_0 L_y)}\,
      \exp(\mathrm{i}nk_0 x)\right],\\
    v(x,y) &=& 2\,\Re\!\left[\sum_{n=1}^{\infty}
      \frac{\hat{f}_n}{nk_0 L_x}\,
      \frac{\sinh(nk_0 y)}{\cosh(nk_0 L_y)}\,
      \exp(\mathrm{i}nk_0 x)\right].
\end{eqnarray}
The wall-normal velocity $v(x,L_y)$ is used as the top boundary condition in the LES. 

\section{SPOD eigenspectra variation across wavenumbers}\label{appB}
\begin{figure}
\centering
\includegraphics{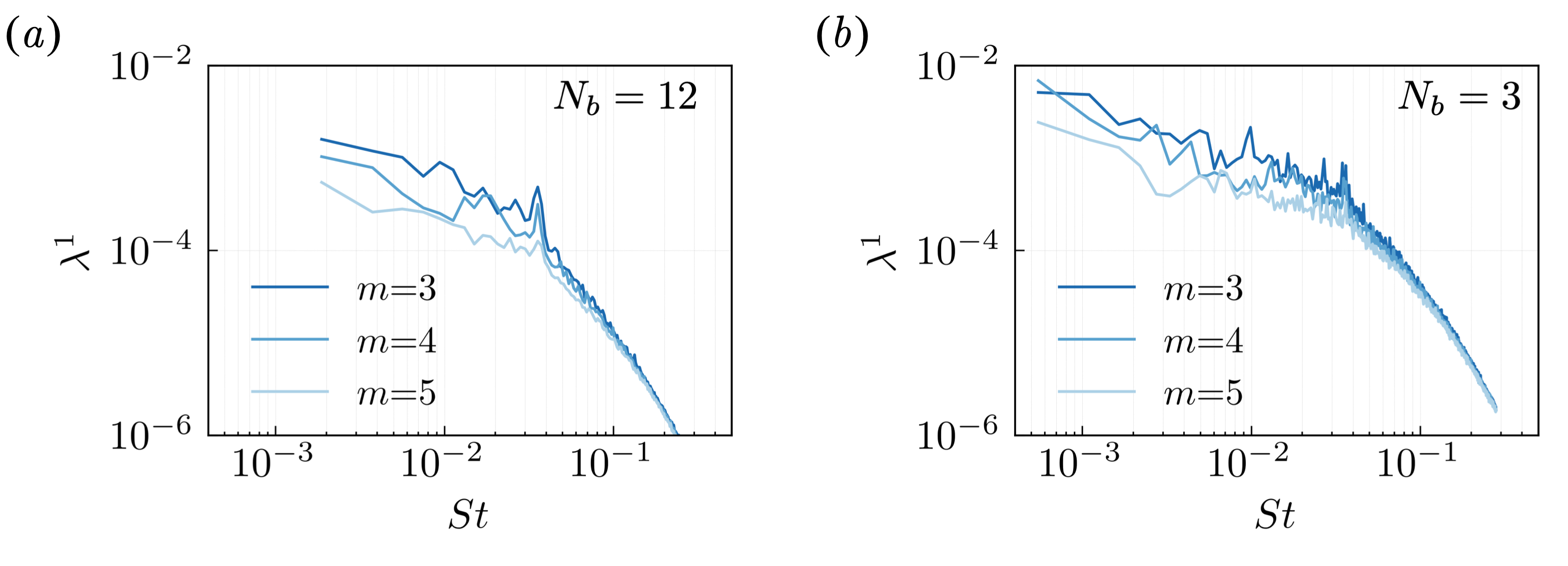}
\caption{\label{fig:appB} The leading SPOD eigenspectra for a range of wavenumber harmonics $3\leq m\leq 5$ obtained using (a) $N_b=3$ and (b)$N_b=12$.}
\end{figure}

In Sec.~\ref{sec:SPOD}, we noted that both the vortex shedding and the low-frequency unsteadiness retain significant energy across a range of spanwise wavenumbers. Fig.~\ref{fig:appB} shows the SPOD eigenspectra for the harmonics $3\leq m\leq 5$, complementing the representative wavenumbers $m=0,1,4$ presented in the main text. Across this range, the leading eigenvalue exhibits distinct peaks near the vortex-shedding frequency ($St=0.036$) and elevated energies in the low-frequency band ($St<0.002$), confirming that these dynamics are distributed over multiple spanwise scales rather than concentrated at a single wavenumber.

\bibliographystyle{unsrtnat}
\bibliography{references}  

@PREAMBLE{
 "\providecommand{\noopsort}[1]{}" 
 # "\providecommand{\singleletter}[1]{#1}%" 
}

@article{michelis2018origin,
  title={On the origin of spanwise vortex deformations in laminar separation bubbles},
  author={Michelis, Theodoros and Yarusevych, Serhiy and Kotsonis, Marios},
  journal={Journal of Fluid Mechanics},
  volume={841},
  pages={81--108},
  year={2018},
  publisher={Cambridge University Press}
}

@inproceedings{sun2014numerical,
  title={Numerical simulations of subsonic and transonic open-cavity flows},
  author={Sun, Yiyang and Nair, Aditya G and Taira, Kunihiko and Cattafesta, Louis N and Bres, Guillaume A and Ukeiley, Lawrence S},
  booktitle={7th AIAA Theoretical Fluid Mechanics Conference},
  pages={3092},
  year={2014}
}

@article{yeh2019resolvent,
  title={Resolvent-analysis-based design of airfoil separation control},
  author={Yeh, Chi-An and Taira, Kunihiko},
  journal={Journal of Fluid Mechanics},
  volume={867},
  pages={572--610},
  year={2019},
  publisher={Cambridge University Press}
}

@article{liu2021unsteady,
  title={Unsteady control of supersonic turbulent cavity flow based on resolvent analysis},
  author={Liu, Qiong and Sun, Yiyang and Yeh, Chi-An and Ukeiley, Lawrence S and Cattafesta, Louis N and Taira, Kunihiko},
  journal={Journal of Fluid Mechanics},
  volume={925},
  year={2021},
  publisher={Cambridge University Press}
}

@article{bres2017unstructured,
  title={Unstructured large-eddy simulations of supersonic jets},
  author={Br{\`e}s, Guillaume A and Ham, Frank E and Nichols, Joseph W and Lele, Sanjiva K},
  journal={AIAA journal},
  volume={55},
  number={4},
  pages={1164--1184},
  year={2017},
  publisher={American Institute of Aeronautics and Astronautics}
}

@article{mckeon2010critical,
  title={A critical-layer framework for turbulent pipe flow},
  author={McKeon, Beverley J and Sharma, Ati S},
  journal={Journal of Fluid Mechanics},
  volume={658},
  pages={336--382},
  year={2010},
  publisher={Cambridge University Press}
}

@article{padovan2020analysis,
  title={Analysis of amplification mechanisms and cross-frequency interactions in nonlinear flows via the harmonic resolvent},
  author={Padovan, Alberto and Otto, Samuel E and Rowley, Clarence W},
  journal={Journal of Fluid Mechanics},
  volume={900},
  pages={A14},
  year={2020},
  publisher={Cambridge University Press}
}

@article{padovan2022analysis,
  title={Analysis of the dynamics of subharmonic flow structures via the harmonic resolvent: Application to vortex pairing in an axisymmetric jet},
  author={Padovan, Alberto and Rowley, Clarence W},
  journal={Physical Review Fluids},
  volume={7},
  number={7},
  pages={073903},
  year={2022},
  publisher={APS}
}

@article{rist2006control,
  title={Control of laminar separation bubbles using instability waves},
  author={Rist, Ulrich and Augustin, Kai},
  journal={AIAA journal},
  volume={44},
  number={10},
  pages={2217--2223},
  year={2006}
}

@article{yarusevych2017steady,
  title={Steady and transient response of a laminar separation bubble to controlled disturbances},
  author={Yarusevych, Serhiy and Kotsonis, Marios},
  journal={Journal of Fluid Mechanics},
  volume={813},
  pages={955--990},
  year={2017},
  publisher={Cambridge University Press}
}

@inproceedings{wu2018response,
  title={Response of a Laminar Separation Bubble to Zero-Net Mass Flux Actuation},
  author={Wu, Wen and Seo, Jung-Hee and Meneveau, Charles and Mittal, Rajat},
  booktitle={2018 Flow Control Conference},
  pages={4018},
  year={2018}
}

@article{marxen2011effect,
  title={The effect of small-amplitude convective disturbances on the size and bursting of a laminar separation bubble},
  author={Marxen, Olaf and Henningson, Dan S},
  journal={Journal of Fluid Mechanics},
  volume={671},
  pages={1--33},
  year={2011},
  publisher={Cambridge University Press}
}

@article{ribeiro2020randomized,
  title={Randomized resolvent analysis},
  author={Ribeiro, Jean H{\'e}lder Marques and Yeh, Chi-An and Taira, Kunihiko},
  journal={Physical Review Fluids},
  volume={5},
  number={3},
  pages={033902},
  year={2020},
  publisher={APS}
}

@article{chu1965energy,
  title={On the energy transfer to small disturbances in fluid flow (Part I)},
  author={Chu, Boa-Teh},
  journal={Acta Mechanica},
  volume={1},
  number={3},
  pages={215--234},
  year={1965},
  publisher={Springer}
}

@article{embacher2014direct,
  title={Direct numerical simulations of laminar separation bubbles: investigation of absolute instability and active flow control of transition to turbulence},
  author={Embacher, Martin and Fasel, HF},
  journal={Journal of fluid mechanics},
  volume={747},
  pages={141--185},
  year={2014},
  publisher={Cambridge University Press}
}

@article{rodriguez2013two,
  title={The two classes of primary modal instability in laminar separation bubbles},
  author={Rodr{\'\i}guez, Daniel and Gennaro, Elmer M and Juniper, Matthew P},
  journal={Journal of Fluid Mechanics},
  volume={734},
  pages={R4},
  year={2013},
  publisher={Cambridge University Press}
}

@article{alam2000direct,
  title={Direct numerical simulation of ‘short’laminar separation bubbles with turbulent reattachment},
  author={Alam, Muhammad and Sandham, Neil D},
  journal={Journal of Fluid Mechanics},
  volume={410},
  pages={1--28},
  year={2000},
  publisher={Cambridge University Press}
}

@article{rist2002investigations,
  title={Investigations of time-growing instabilities in laminar separation bubbles},
  author={Rist, Ulrich and Maucher, Ulrich},
  journal={European Journal of Mechanics-B/Fluids},
  volume={21},
  number={5},
  pages={495--509},
  year={2002},
  publisher={Elsevier}
}

@article{rodriguez2010structural,
  title={Structural changes of laminar separation bubbles induced by global linear instability},
  author={Rodriguez, Daniel and Theofilis, Vassilis},
  journal={Journal of Fluid Mechanics},
  volume={655},
  pages={280--305},
  year={2010},
  publisher={Cambridge University Press}
}

@book{gaster1967structure,
  title={The structure and behaviour of laminar separation bubbles},
  author={Gaster, Michael},
  year={1967},
  publisher={Citeseer}
}

@article{balzer2016numerical,
  title={Numerical investigation of the role of free-stream turbulence in boundary-layer separation},
  author={Balzer, Wolfgang and Fasel, Hermann F},
  journal={Journal of Fluid Mechanics},
  volume={801},
  pages={289--321},
  year={2016},
  publisher={Cambridge University Press}
}

@article{islam2024identification,
  title={Identification of cross-frequency interactions in compressible cavity flow using harmonic resolvent analysis},
  author={Islam, Md Rashidul and Sun, Yiyang},
  journal={Journal of Fluid Mechanics},
  volume={1000},
  pages={A13},
  year={2024},
  publisher={Cambridge University Press}
}

@techreport{balay2019petsc,
  title={PETSc users manual},
  author={Balay, Satish and Abhyankar, Shrirang and Adams, Mark and Brown, Jed and Brune, Peter and Buschelman, Kris and Dalcin, Lisandro and Dener, Alp and Eijkhout, Victor and Gropp, William and others},
  year={2019},
  number={ANL-95/11- Revision 3.12},
  institution={Argonne National Laboratory}
}

@article{hernandez2005slepc,
  title={SLEPc: A scalable and flexible toolkit for the solution of eigenvalue problems},
  author={Hernandez, Vicente and Roman, Jose E and Vidal, Vicente},
  journal={ACM Transactions on Mathematical Software (TOMS)},
  volume={31},
  number={3},
  pages={351--362},
  year={2005},
  publisher={ACM New York, NY, USA}
}

@inproceedings{amestoy2000mumps,
  title={MUMPS: a general purpose distributed memory sparse solver},
  author={Amestoy, Patrick R and Duff, Iain S and L’Excellent, Jean-Yves and Koster, Jacko},
  booktitle={International Workshop on Applied Parallel Computing},
  pages={121--130},
  year={2000},
  organization={Springer}
}

@article{vreman2004eddy,
  title={An eddy-viscosity subgrid-scale model for turbulent shear flow: Algebraic theory and applications},
  author={Vreman, AW},
  journal={Physics of fluids},
  volume={16},
  number={10},
  pages={3670--3681},
  year={2004},
  publisher={American Institute of Physics}
}

@article{borgmann2025experimental,
  title={Experimental and numerical investigations of transition in a pressure-gradient-induced laminar separation bubble},
  author={Borgmann, David and Hosseinverdi, Shirzad and Little, Jesse and Fasel, Hermann},
  journal={Journal of Fluid Mechanics},
  volume={1007},
  pages={A23},
  year={2025},
  publisher={Cambridge University Press}
}

@article{sun2020resolvent,
  title={Resolvent analysis of compressible laminar and turbulent cavity flows},
  author={Sun, Yiyang and Liu, Qiong and Cattafesta III, Louis N and Ukeiley, Lawrence S and Taira, Kunihiko},
  journal={AIAA journal},
  volume={58},
  number={3},
  pages={1046--1055},
  year={2020},
  publisher={American Institute of Aeronautics and Astronautics}
}

@article{spalart2000mechanisms,
  title={Mechanisms of transition and heat transfer in a separation bubble},
  author={Spalart, Philippe R and Strelets, Michael Kh},
  journal={Journal of Fluid Mechanics},
  volume={403},
  pages={329--349},
  year={2000},
  publisher={Cambridge University Press}
}

@article{welch1967use,
  title={The use of fast Fourier transform for the estimation of power spectra: A method based on time averaging over short, modified periodograms},
  author={Welch, Peter},
  journal={IEEE Transactions on audio and electroacoustics},
  volume={15},
  number={2},
  pages={70--73},
  year={1967},
  publisher={IEEE}
}

@article{zaman1989natural,
  title={A natural low-frequency oscillation of the flow over an airfoil near stalling conditions},
  author={Zaman, KBMQ and McKinzie, DJ and Rumsey, CL},
  journal={Journal of Fluid Mechanics},
  volume={202},
  pages={403--442},
  year={1989},
  publisher={Cambridge University Press}
}

@article{malmir2024low,
  title={Low-frequency unsteadiness in laminar separation bubbles},
  author={Malmir, Fatemeh and Di Labbio, Giuseppe and Le Floc'h, Arnaud and Dufresne, Louis and Weiss, Julien and V{\'e}tel, J{\'e}r{\^o}me},
  journal={Journal of Fluid Mechanics},
  volume={999},
  pages={A99},
  year={2024},
  publisher={Cambridge University Press}
}

@article{michelis2017response,
  title={Response of a laminar separation bubble to impulsive forcing},
  author={Michelis, Theodoros and Yarusevych, Serhiy and Kotsonis, Marios},
  journal={Journal of Fluid Mechanics},
  volume={820},
  pages={633--666},
  year={2017},
  publisher={Cambridge University Press}
}

@article{rodriguez2021self,
  title={Self-excited primary and secondary instability of laminar separation bubbles},
  author={Rodr{\'\i}guez, Daniel and Gennaro, Elmer M and Souza, Leandro F},
  journal={Journal of Fluid Mechanics},
  volume={906},
  pages={A13},
  year={2021},
  publisher={Cambridge University Press}
}

@book{lumley2012stochastic,
  title={Stochastic tools in turbulence},
  author={Lumley, John L},
  year={2012},
  publisher={Elsevier}
}

@article{towne2018spectral,
  title={Spectral proper orthogonal decomposition and its relationship to dynamic mode decomposition and resolvent analysis},
  author={Towne, Aaron and Schmidt, Oliver T and Colonius, Tim},
  journal={Journal of Fluid Mechanics},
  volume={847},
  pages={821--867},
  year={2018},
  publisher={Cambridge University Press}
}

@article{sirovich1987turbulence,
  title={Turbulence and the dynamics of coherent structures. I. Coherent structures},
  author={Sirovich, Lawrence},
  journal={Quarterly of applied mathematics},
  volume={45},
  number={3},
  pages={561--571},
  year={1987}
}

@article{rolandi2024invitation,
  title={An invitation to resolvent analysis},
  author={Rolandi, Laura Victoria and Ribeiro, Jean H{\'e}lder Marques and Yeh, Chi-An and Taira, Kunihiko},
  journal={Theoretical and Computational Fluid Dynamics},
  volume={38},
  number={5},
  pages={603--639},
  year={2024},
  publisher={Springer}
}

@article{beneddine2016conditions,
  title={Conditions for validity of mean flow stability analysis},
  author={Beneddine, Samir and Sipp, Denis and Arnault, Anthony and Dandois, Julien and Lesshafft, Lutz},
  journal={Journal of Fluid Mechanics},
  volume={798},
  pages={485--504},
  year={2016},
  publisher={Cambridge University Press}
}

@book{jovanovic2004modeling,
  title={Modeling, analysis, and control of spatially distributed systems},
  author={Jovanovic, Mihailo R},
  year={2004},
  publisher={University of California, Santa Barbara}
}

@article{yeh2020resolvent,
  title = {Resolvent analysis of an airfoil laminar separation bubble at $\text{Re}=500\phantom{\rule{0.16em}{0ex}}000$},
  author = {Yeh, Chi-An and Benton, Stuart I. and Taira, Kunihiko and Garmann, Daniel J.},
  journal = {Phys. Rev. Fluids},
  volume = {5},
  issue = {8},
  pages = {083906},
  numpages = {24},
  year = {2020},
  month = {Aug},
  publisher = {American Physical Society},
  doi = {10.1103/PhysRevFluids.5.083906},
  url = {https://link.aps.org/doi/10.1103/PhysRevFluids.5.083906}
}

@article{rolandi2025biglobal,
  title={Biglobal resolvent analysis of separated flow over a NACA0012 airfoil},
  author={Rolandi, Laura Victoria and Smith, Luke and Amitay, Michael and Theofilis, Vassilis and Taira, Kunihiko},
  journal={Journal of Fluid Mechanics},
  volume={1021},
  pages={A53},
  year={2025},
  publisher={Cambridge University Press}
}

@article{tani1964low,
  title={Low-speed flows involving bubble separations},
  author={Tani, Itiro},
  journal={Progress in Aerospace Sciences},
  volume={5},
  pages={70--103},
  year={1964},
  publisher={Elsevier}
}

@article{visbal2023passive,
  title={Passive control of dynamic stall using a flow-driven micro-cavity actuator: MR Visbal, DJ Garmann},
  author={Visbal, Miguel R and Garmann, Daniel J},
  journal={Theoretical and Computational Fluid Dynamics},
  volume={37},
  number={3},
  pages={289--303},
  year={2023},
  publisher={Springer}
}

@article{visbal2018exploration,
  title={Exploration of high-frequency control of dynamic stall using large-eddy simulations},
  author={Visbal, Miguel R and Benton, Stuart I},
  journal={AIAA Journal},
  volume={56},
  number={8},
  pages={2974--2991},
  year={2018},
  publisher={American Institute of Aeronautics and Astronautics}
}

@article{marxen2010mean,
  title={Mean flow deformation in a laminar separation bubble: separation and stability characteristics},
  author={Marxen, Olaf and Rist, Ulrich},
  journal={Journal of Fluid Mechanics},
  volume={660},
  pages={37--54},
  year={2010},
  publisher={Cambridge University Press}
}

@article{freund1997proposed,
  title={Proposed inflow/outflow boundary condition for direct computation of aerodynamic sound},
  author={Freund, Jonathan B},
  journal={AIAA journal},
  volume={35},
  number={4},
  pages={740--742},
  year={1997}
}

@article{leclercq2023mean,
  title={Mean resolvent operator of a statistically steady flow},
  author={Leclercq, Colin and Sipp, Denis},
  journal={Journal of Fluid Mechanics},
  volume={968},
  pages={A13},
  year={2023},
  publisher={Cambridge University Press}
}

@article{jones2008direct,
  title={Direct numerical simulations of forced and unforced separation bubbles on an airfoil at incidence},
  author={Jones, LE and Sandberg, Richard D and Sandham, Neil D},
  journal={Journal of Fluid Mechanics},
  volume={602},
  pages={175--207},
  year={2008},
  publisher={Cambridge University Press}
}

@article{kurelek2016coherent,
  title={Coherent structures in the transition process of a laminar separation bubble},
  author={Kurelek, John W and Lambert, Andrew R and Yarusevych, Serhiy},
  journal={AIAA Journal},
  volume={54},
  number={8},
  pages={2295--2309},
  year={2016},
  publisher={American Institute of Aeronautics and Astronautics}
}

@article{hosseinverdi2019numerical,
  title={Numerical investigation of laminar--turbulent transition in laminar separation bubbles: the effect of free-stream turbulence},
  author={Hosseinverdi, Shirzad and Fasel, Hermann F},
  journal={Journal of Fluid Mechanics},
  volume={858},
  pages={714--759},
  year={2019},
  publisher={Cambridge University Press}
}

@article{ran2019stochastic,
  title={Stochastic receptivity analysis of boundary layer flow},
  author={Ran, Wei and Zare, Armin and Hack, MJ Philipp and Jovanovi{\'c}, Mihailo R},
  journal={Physical Review Fluids},
  volume={4},
  number={9},
  pages={093901},
  year={2019},
  publisher={APS}
}

@article{cherubini2010onset,
  title={The onset of three-dimensional centrifugal global modes and their nonlinear development in a recirculating flow over a flat surface},
  author={Cherubini, Stefania and Robinet, J-Ch and De Palma, Pietro and Alizard, Fr{\'e}d{\'e}ric},
  journal={Physics of Fluids},
  volume={22},
  number={11},
  year={2010},
  publisher={AIP Publishing}
}

@article{cura2025linear,
  title={Linear modeling of a family of turbulent separation bubbles},
  author={Cura, Carolina and Hanifi, Ardeshir and Cavalieri, Andr{\`e} VG and Weiss, Julien},
  journal={Physical Review Fluids},
  volume={10},
  number={11},
  pages={114607},
  year={2025},
  publisher={APS}
}

@article{thakor2024responses,
  title={Responses to disturbance of supersonic shear layer: Input-output analysis},
  author={Thakor, Mitesh and Sun, Yiyang and Gaitonde, Datta V},
  journal={Physical Review Fluids},
  volume={9},
  number={8},
  pages={084603},
  year={2024},
  publisher={APS}
}

@article{marxen2003combined,
  title={A combined experimental/numerical study of unsteady phenomena in a laminar separation bubble},
  author={Marxen, O and Lang, M and Rist, Ulrich and Wagner, Siegfried},
  journal={Flow, Turbulence and Combustion},
  volume={71},
  number={1},
  pages={133--146},
  year={2003},
  publisher={Springer}
}

@article{theofilis2000origins,
  title={On the origins of unsteadiness and three-dimensionality in a laminar separation bubble},
  author={Theofilis, Vassilios and Hein, Stefan and Dallmann, Uwe},
  journal={Philosophical Transactions of the Royal Society of London. Series A: Mathematical, Physical and Engineering Sciences},
  volume={358},
  number={1777},
  pages={3229--3246},
  year={2000},
  publisher={The Royal Society}
}

@article{marxen2004effect,
  title={Effect of spanwise-modulated disturbances on transition in a separated boundary layer},
  author={Marxen, Olaf and Rist, Ulrich and Wagner, Siegfried},
  journal={AIAA journal},
  volume={42},
  number={5},
  pages={937--944},
  year={2004}
}

@article{collis2004issues,
  title={Issues in active flow control: theory, control, simulation, and experiment},
  author={Collis, S Scott and Joslin, Ronald D and Seifert, Avi and Theofilis, Vassilis},
  journal={Progress in aerospace sciences},
  volume={40},
  number={4-5},
  pages={237--289},
  year={2004},
  publisher={Elsevier}
}

@article{borgmann2025active,
  title={Active control of transition to turbulence in laminar separation bubbles},
  author={Borgmann, David and Little, Jesse and Fasel, Hermann},
  journal={Journal of Fluid Mechanics},
  volume={1016},
  pages={A59},
  year={2025},
  publisher={Cambridge University Press}
}

@article{kurelek2023superposition,
  title={Superposition of AC-DBD plasma actuator outputs for three-dimensional disturbance production in shear flows},
  author={Kurelek, John W and Kotsonis, Marios and Yarusevych, Serhiy},
  journal={Experiments in Fluids},
  volume={64},
  number={4},
  pages={84},
  year={2023},
  publisher={Springer}
}

@article{toppings2026bursting,
  title={Bursting of a laminar separation bubble subject to periodic forcing on a pitching airfoil},
  author={Toppings, Connor and Michelis, Theodoros and Kotsonis, Marios and Yarusevych, Serhiy},
  journal={Physical Review Fluids},
  volume={11},
  number={7},
  pages={073901},
  year={2026},
  publisher={APS}
}

@article{yarusevych2017effect,
  title={Effect of local DBD plasma actuation on transition in a laminar separation bubble},
  author={Yarusevych, Serhiy and Kotsonis, Marios},
  journal={Flow, Turbulence and Combustion},
  volume={98},
  number={1},
  pages={195--216},
  year={2017},
  publisher={Springer}
}

@article{diwan2009origin,
  title={On the origin of the inflectional instability of a laminar separation bubble},
  author={Diwan, Sourabh S and Ramesh, ON},
  journal={Journal of Fluid Mechanics},
  volume={629},
  pages={263--298},
  year={2009},
  publisher={Cambridge University Press}
}

@incollection{colonius2025modal,
  title={Modal decomposition},
  author={Colonius, Tim and Towne, Aaron},
  booktitle={Data Driven Analysis and Modeling of Turbulent Flows},
  pages={27--81},
  year={2025},
  publisher={Elsevier}
}

@article{sun2017biglobal,
  title={Biglobal instabilities of compressible open-cavity flows},
  author={Sun, Yiyang and Taira, Kunihiko and Cattafesta, Louis N and Ukeiley, Lawrence S},
  journal={Journal of Fluid Mechanics},
  volume={826},
  pages={270--301},
  year={2017},
  publisher={Cambridge University Press}
}

@book{sun2017global,
  title={Global Stability Analysis and Control of Compressible Flows over Rectangular Cavities},
  author={Sun, Yiyang},
  year={2017},
  publisher={The Florida State University}
}

@article{yeung2024high,
  title={High-speed microactuation in a supersonic dual-stream jet flow},
  author={Yeung, Melissa and Gaitonde, Datta V and Sun, Yiyang},
  journal={Journal of Fluid Mechanics},
  volume={998},
  pages={A53},
  year={2024},
  publisher={Cambridge University Press}
}

@article{lugrin2021transition,
  title={Transition scenario in hypersonic axisymmetrical compression ramp flow},
  author={Lugrin, Mathieu and Beneddine, Samir and Leclercq, Colin and Garnier, Eric and Bur, Reynald},
  journal={Journal of Fluid Mechanics},
  volume={907},
  pages={A6},
  year={2021},
  publisher={Cambridge University Press}
}

@article{farghadan2026wave, 
title={Wave interactions in a screeching jet}, volume={1040}, 
DOI={10.1017/jfm.2026.11818}, 
journal={Journal of Fluid Mechanics}, 
author={Farghadan, Ali and Beekman, Jayson and Nogueira, Petronio and Edgington-Mitchell, Daniel and Towne, Aaron}, 
year={2026}, 
pages={A40}}

@article{linot2025extracting,
  title={Extracting dominant dynamics about unsteady base flows},
  author={Linot, Alec J and Lopez-Doriga, Barbara and Zhong, Yonghong and Taira, Kunihiko},
  journal={Fluid Dynamics Research},
  volume={57},
  number={3},
  pages={031401},
  year={2025},
  publisher={IOP Publishing}
}

@article{esfahani2019flow,
  title={Flow separation control over a thin post-stall airfoil: Effects of excitation frequency},
  author={Esfahani, Ata and Webb, Nathan and Samimy, Mo},
  journal={AIAA Journal},
  volume={57},
  number={5},
  pages={1826--1838},
  year={2019},
  publisher={American Institute of Aeronautics and Astronautics}
}

@article{islam2025effect,
  title={Effect of cavity-induced perturbation interactions on the transitional flow after the trailing edge},
  author={Islam, Md Rashidul and Sun, Yiyang},
  journal={arXiv preprint arXiv:2511.12317},
  year={2025}
}

\end{document}